\documentclass[aps,prx,cha,graphicx,amsmath,amssymb,reprint,onecolumn,
superscriptaddress,floatfix]{revtex4}
\usepackage[]{graphicx}

\begin{document}
\title{A design framework for actively crosslinked filament networks}
\author{Sebastian F\"urthauer}
\affiliation{Center for Computational Biology, Flatiron Institute, New York, NY 10010, USA} 
\author{Daniel J. Needleman}
\affiliation{Paulson School of Engineering \& Applied Science and Department of
Molecular \& Cellular Biology, Harvard University, Cambridge, MA 02138, USA}
\author{Michael J. Shelley}
\affiliation{Center for Computational Biology, Flatiron Institute, New York, NY 10010, USA}
\affiliation{Courant Institute, New York University, New York, NY 10012, USA}

\begin{abstract}
    Living matter moves, deforms, and organizes itself. In cells this is made
    possible by networks of polymer filaments and crosslinking molecules that
    connect filaments to each other and that act as motors to do mechanical work on
    the network.  For the case of highly cross-linked filament networks, we
    discuss how the material properties of assemblies emerge from the forces
    exerted by microscopic agents.  First, we introduce a phenomenological model
    that characterizes the forces that crosslink populations exert between
    filaments.  Second, we derive a theory that predicts the material properties
    of highly crosslinked filament networks, given the crosslinks present.
    Third, we discuss which properties of crosslinks set the material properties
    and behavior of highly crosslinked cytoskeletal networks.  The work
    presented here, will enable the better understanding of cytoskeletal
    mechanics and its molecular underpinnings. This theory is also a first step
    towards a theory of how molecular perturbations impact cytoskeletal
    organization, and provides a framework for designing cytoskeletal networks
    with desirable properties in the lab.  
\end{abstract}

\maketitle

\section{Introduction}
Materials made from constituents that use energy to move are called active.
These inherently out of equilibrium systems have attractive physical properties:
active materials can spontaneously form patterns \cite{bois2011pattern},
collectively move \cite{voituriez2005spontaneous, furthauer2012taylor,
wioland2013confinement}, self-organize into structures
\cite{salbreux2009hydrodynamics,brugues2014physical}, and do work
\cite{marchetti2013hydrodynamics}.  Biology, through evolution, has found ways
to exploit this potential. The cytoskeleton, an active material made from
biopolymer filaments and molecular scale motors, drives cellular functions with
remarkable spatial and temporal coordination \cite{alberts_2002_book,
howard2001mechanics}. The ability of cells to move, divide, and deform relies on
this robust, dynamic and adaptive material. To understand the molecular
underpinnings of cellular mechanics and design similarly useful active matter
systems in the lab, a theory that predicts their behavior from the interactions
between their constituents is needed. The aim of this paper, is to address this
challenge for highly crosslinked systems made from rigid rod-like filaments and 
molecular scale motors.

The large-scale physics of active materials can be described by phenomenological
theories, which are derived from symmetry considerations and conservation laws,
without making assumptions on the detailed molecular scale interactions that
give rise to the materials properties
\cite{kruse2004asters, furthauer2012active, julicher2018hydrodynamic}. This has
allowed exploring the exotic properties of active materials, and the
quantitative description of subcellular structures, such as the spindle
\cite{brugues2014physical, oriola2020active} (the structure that segregates
chromosomes during cell division) and the cell
cortex\cite{mayer2010anisotropies,salbreux2012actin,naganathan2014active,naganathan2018morphogenetic}
(the structure that provides eukaryotic cells with the ability to control their
shape), even though the microscale processes at work often remain opaque. In
contrast, understanding how molecular perturbations affect cellular scale
structures requires theories that explain how material properties depend on
the underlying molecular behaviors. Designing active materials with desirable properties in the lab will
also require the ability to predict how emergent properties of materials result
from their constituents
\cite{foster2019cytoskeletal}. Until now, the attempts to bridge this gap have
relied heavily on computational methods \cite{belmonte2017theory,
foster2017connecting, gao2015multiscalepolar}, or were restricted to sparsely
crosslinked systems
\cite{liverpool2003instabilities, liverpool2005bridging,
    liverpool2008hydrodynamics, aranson2005pattern, 
    saintillan2008instabilitiesPRL}, one dimensional
    systems \cite{kruse2000actively, kruse2003self}, or systems with
permanent crosslinks \cite{broedersz2014modeling}. Our interest here are cytoskeletal networks,
which are in general highly crosslinked by tens to hundreds of transient
crosslinks linking each filament into the network. In this regime, 
the forces generated by different crosslinks in the network balance against each
other, and not against friction with the surrounding medium, as they would
in a sparsely crosslinked regime \cite{furthauer2019self}.

We derive how the large scale properties of an actively crosslinked network of
cytoskeletal filaments depend on the micro-scale interactions between its
components. This theory generalizes our earlier work on one specific type of
motor-filament mixture, XCTK2 and microtubules
\cite{furthauer2019self, striebel2020mechanistic}, by introducing a generic
phenomenological model to describe the forces that crosslink populations exert
between filaments.

The structure of this paper is as follows. In section \ref{sse:generic}, we
discuss the force and torque balance for systems of interacting particles, and
specialize to the case of interacting rod-like filaments. This will allow us to
introduce key concepts of the continuum description, such as the network stress
tensor. Next, in section \ref{sse:interactions}, we present a phenomenological
model for crosslink interactions between filaments, that can describe the
properties of many different types of crosslinks in terms of just a few
parameters, which we call crosslink moments. In section \ref{sse:continuum}, we
derive the continuum theory for highly crosslinked active networks and obtain
the equations of motion for these systems.  Finally, in section \ref{sse:macro}
we give an overview of the main predictions of our theory and discuss the
consequences of specific micro-scale properties for the mechanical properties of
the consequent active material. We summarize and contextualize our findings in
the discussion section \ref{sse:discussion}.

\section{Force and torque balance in systems of interacting rod-like
particles}\label{sse:generic} 
We start by discussing the generic framework of our description. In this section
we give equations for particle, momentum and angular momentum conservation and
introduce the stress tensor, for generic systems of particles with short ranged
interactions. We then specialize to the case of interacting rod-like filaments,
which form the networks that we study here.

\subsection{Particle Number Continuity}
Consider a material that consists of a large number $N$ of particles,
that are characterized by their center of mass positions $\mathbf x_i$ and their
orientations $\mathbf p_i$, where $|\mathbf p_i| = 1$ is an unit vector and 
$i$ is the particle index. 
We define the particle number density
\begin{equation}
    \psi(\mathbf x, \mathbf p) = \sum_i \delta(\mathbf x - \mathbf
    x_i)\delta(\mathbf p - \mathbf p_i).
    \label{eq:density_definition}
\end{equation}
Here and in the following $\delta(\mathbf x-\mathbf x_i)$ has dimensions of
inverse volume, while $\delta(\mathbf p-\mathbf p_i)$ is dimensionless.
Ultimately, our goal is to predict how $\psi$ changes over time.  This is given
by the Smoluchowski equation
\begin{equation}
    \partial_t \psi(\mathbf x, \mathbf p) = 
    -\nabla\cdot\left(\dot{\mathbf x} \psi \right)
    -\partial_\mathbf{p}\cdot\left(\dot{\mathbf p} \psi \right),
    \label{eq:smoluchowski}
\end{equation}
where 
\begin{equation}
    \dot{\mathbf x}\psi =\sum_i \dot{\mathbf x}_i\delta(\mathbf x - \mathbf
x_i)\delta(\mathbf p - \mathbf p_i)
\end{equation}
and
\begin{equation}
\dot{\mathbf p}\psi =\sum_i \dot{\mathbf p}_i\delta(\mathbf x - \mathbf
x_i)\delta(\mathbf p - \mathbf p_i)
\end{equation}
define $\dot{\mathbf x}$ and $\dot{\mathbf p}$, the fluxes of particle
position and orientation. The aim of this paper is to derive $\dot{\mathbf x}$
and $\dot{\mathbf p}$, from the forces and torques that
act on and between particles. 
\subsection{Force Balance}
Each particle in the active network obeys Newton's laws of motion.
That is
\begin{equation}
    \dot{\mathbf g}_i = \sum_j \mathbf F_{ij} \mathbf
     + \mathbf F_i^\mathrm{(drag)},
\label{eq:Newton_single_particle}
\end{equation}
where $\mathbf g_i$ is the particle momentum, and $\mathbf F_{ij}$ is the force
that particle $j$ exerts on particle $i$. Moreover, $\mathbf
F_i^\mathrm{(drag)}$ is the drag force between the particle $i$ and the fluid in
which it is immersed. Momentum conservation
implies $\mathbf F_{ij}
= - \mathbf F_{ji}$. We are interested in systems where the
direct particle-particle interactions are short ranged. This means that
$\mathbf F_{ij}\ne 0$ only if $|\mathbf x_i -\mathbf x_j|<d$, where $d$ is an
interaction length that is small (relative to system size).  

The momentum density is defined by
\begin{equation} 
    \mathbf g = \sum_i \delta(\mathbf x - \mathbf x_i) \mathbf g_i 
\end{equation}
which, using Eq.~{(\ref{eq:Newton_single_particle})}, obeys
\begin{eqnarray}
\partial_t \mathbf g &+& \nabla\cdot\sum_i\delta(\mathbf x -
\mathbf x_i)\mathbf v_i  \mathbf
g_i 
= \sum_{i,j}\delta(\mathbf x - \mathbf x_i)
\mathbf F_{ij} 
+ \sum_{i}\delta(\mathbf x - \mathbf x_i)
\mathbf F^\mathrm{(drag)}_{i} ,
\nonumber\\
\label{eq:Fb_continuum}
\end{eqnarray}
where $\mathbf v_i = \dot{\mathbf x}_i$ is the velocity of the $i$-th particle. 
The terms on the left hand side of Eq.~{(\ref{eq:Fb_continuum})} are inertial,
and in the overdamped limit, relevant to the systems studied here, they are
vanishingly small. 
Interactions between particles are described by the first term on the
right hand side of Eq.~{(\ref{eq:Fb_continuum})} and generate a momentum density flux
$\mathbf\Sigma$ (the stress tensor) through the material. To wit, using that $d$
is small, so that particle-particle interactions are short-ranged, gives
\begin{eqnarray}
    &~&\sum_{i,j}\delta(\mathbf x - \mathbf x_i)\mathbf F_{ij}
=\frac{1}2\sum_{i,j}\left(
\delta(\mathbf x - \mathbf x_i)
- \delta(\mathbf x - \mathbf x_j)\right)\mathbf F_{ij}
\nonumber\\
&=&-\nabla \cdot \sum_{i,j}
\delta\left(\mathbf x - \mathbf x_i\right)\frac{\mathbf x_i -\mathbf x_j}2
\mathbf F_{ij} +\mathcal{O}(d^3)  
\nonumber \\ 
&=&\nabla\cdot\mathbf\Sigma.
\end{eqnarray}
where
\begin{equation}
\mathbf\Sigma = -\sum_{i,j} \delta\left(\mathbf x - \mathbf
x_i \right)\frac{\mathbf x_{i} - \mathbf x_{j}}2\mathbf F_{ij}
+\mathcal{O}(d^3). 
\label{eq:define_stress}
\end{equation}
Note that Eq.~{(\ref{eq:define_stress})} does not necessarily
produce a symmetric stress tensor. Force couples for which $\mathbf
F_{ij}$ and $\mathbf x_{i}-\mathbf x_{j}$ are not parallel generate antisymmetric stress
contributions, since these couples are not torque free. We discuss how to reconcile this
with angular momentum conservation in Appendix \ref{app:angular}. 
The drag force density is
\begin{equation}
    \mathbf f = \sum_i\delta(\mathbf x-\mathbf x_i)\mathbf
    F_i^\mathrm{(drag)},
    \label{eq:define_permaeation_force}
\end{equation}
and after dropping inertial terms, the force balance reads
\begin{equation}
    \nabla\cdot\mathbf\Sigma + \mathbf f = \mathbf 0,
    \label{eq:gel_force_balance}
\end{equation}
and the total force on particle $i$ obeys
\begin{equation}
    \sum_j \mathbf F_{ij} + \mathbf F_{i}^\mathrm{(drag)} = \mathbf 0.
    \label{eq:Force_balance_mti}
\end{equation}
This completes the discussion of the force balance of the
system. We next discuss angular momentum conservation.

\subsection{Torque Balance}
\label{sse:torque_generic}
The total angular momentum of
particle $i$,
\begin{equation}
\mathbf \ell^\mathrm{(tot)}_i = \mathbf \ell_i +\mathbf
x_i\times \mathbf g_i,
\label{eq:total_angular_momentum_i}
\end{equation}
is conserved, where $\mathbf \ell_i$ is its spin angular momentum and its
$\mathbf x_i\times\mathbf g_i$ its orbital angular momentum.
Newton's laws imply that 
\begin{equation}
    \dot{\mathbf\ell}_i = \sum_j\mathbf T_{ij} + \mathbf  T_{i}^\mathrm{(drag)},
\end{equation}
where $\mathbf T_{ij}$ is the torque exerted by particle $j$ on particle $i$, in the
frame of reference moving with particle $i$,nd
$\mathbf T_i^\mathrm{(drag)}$ is the torque from interaction with the medium, in the
same frame of reference. Importantly, since the total angular momentum is a
conserved quantity, the total torque transmitted between particles
$\mathbf  T_{ij} + \mathbf x_i\times \mathbf F_{ij} = -\mathbf  T_{ji}  -
\mathbf x_j\times \mathbf F_{ji}$ is odd upon exchange of the particle indices
$i$ and $j$.
Taking a time derivative of Eq.{~(\ref{eq:total_angular_momentum_i})} and using
Eq.{~(\ref{eq:Newton_single_particle})} leads to the torque balance equation for particle $i$
\begin{equation}
\sum_j
\left( \mathbf T_{ij} + \mathbf x_i\times \mathbf F_{ij} \right) + \mathbf T_i^\mathrm{(drag)}
+\mathbf x_i \times \mathbf  F_{i}^\mathrm{(drag)} = \mathbf 0,
\end{equation}
and thus
\begin{equation}
\sum_j\mathbf T_{ij}  + \mathbf T_i^\mathrm{(drag)} = \mathbf 0,
\end{equation}
where we ignored the inertial term $\mathbf v_i \times \mathbf g_i$ and used Eq.~{(\ref{eq:Force_balance_mti})}.
 The angular momentum fluxes associated with spin, orbital and total
angular momentum are discussed in Appendix{~\ref{app:angular}} for completeness. 

\subsection{The special case of rod-like filaments}
\begin{figure}[h]
    \centering
    \includegraphics[width=0.6\columnwidth]{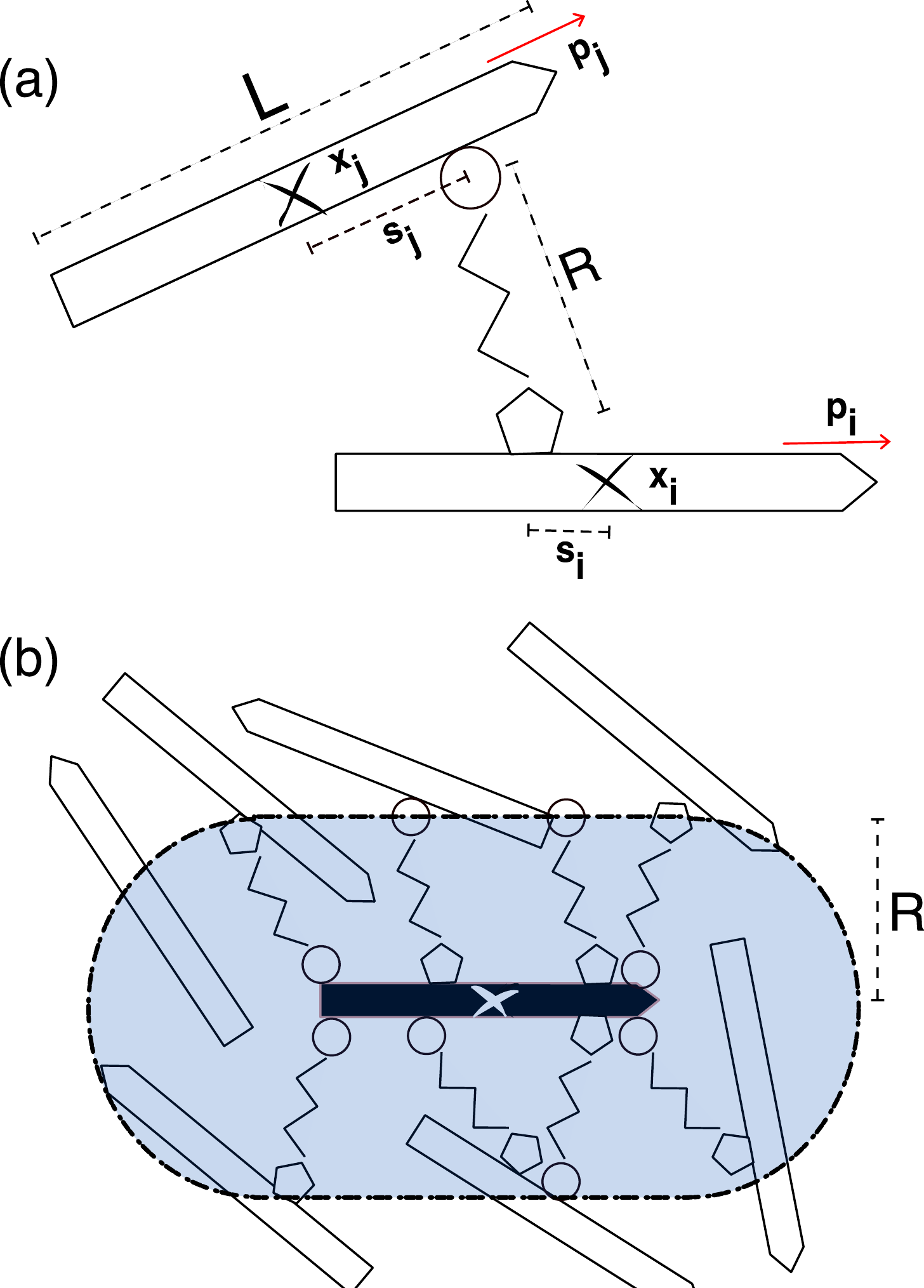}
    \caption{
        a/ Interaction between two cytoskeletal filament $i$ and $j$ via a molecular
        motor.  Filaments are characterized by their positions $\mathbf x_i,
        \mathbf x_j$, their
        orientations $\mathbf p_i, \mathbf p_j$, and connect by a motor between
        arc-length position $s_i, s_j$. A motor consist of two
        heads that can be different (circle, pentagon) and are connected by a
        linker (black zig-zag) of lengt $R$
        b/ The total force on filament $i$ is given by the sum of the forces exerted by
    all $a$ (circle) and $b$ (pentagon) heads, which connect the filament into the
    network. The shaded area shows all geometrically accessible positions that
    can be crosslinked to the central (black) filament. } 
\label{fig:cartoons}
\end{figure}
We now specialize to rod-like particles, such as the microtubules and actin filaments that
make up the cytoskeleton. In particular, we calculate the objects $\mathbf
F_{ij}$, $\mathbf T_{ij}$, and $\mathbf
\Sigma$ from prescribed interaction forces and torques along
rod-like particles. 
\subsubsection{Forces}
Again, filament $i$ is described by it center of mass $\mathbf{x}_i$ and orientation vector
$\mathbf{p_i}$. All filaments are taken as having the same length $L$, and position along filament $i$ is given by $\mathbf{x}_i+s_i\mathbf{p}_i$, where $s_i\in[-L/2,L/2]$ is the
signed arclength. 
We consider the vectorial momentum flux from arclength position $s_i$ on filament $i$
to arclength position $s_j$ on filament $j$ 
\begin{equation}
    \mathbf f_{ij} = \mathbf f_{ij}\left(s_i, s_j \right).
\end{equation}
where $\mathbf f_{ij} = - \mathbf f_{ji}$ and having dimensions of force over
area, i.e. a stress.  
Here we focus on forces generated by crosslinks;
see Fig.~{\ref{fig:cartoons}} (a).    
The total force between two particles is 
\begin{eqnarray}
    \mathbf F_{ij} &=& 
    \left\lfloor \delta(\mathbf x -
    \mathbf x_j - s_j \mathbf p_j) \mathbf f_{ij}\right\rceil^{ij}_{\Omega(\mathbf x_i
    +s_i\mathbf p_i)},
    \nonumber\\
\end{eqnarray}
where the brackets $\lfloor\cdots\rceil^{ij}_{\Omega(\mathbf x_i)}$
denote the operation
\begin{equation}
    \lfloor\phi\rceil_{\Omega(\mathbf x_i)}^{ij} = \int\limits_{-\frac{L}{2}}^{\frac{L}{2}} ds_i
    \int\limits_{-\frac{L}{2}}^{\frac{L}{2}} ds_j
    \int\limits_{\Omega(\mathbf x_i)}d\mathbf x^3 \phi, 
    \label{eq:bracket_def}
\end{equation}
where $\phi$ is a dummy argument and $\Omega$ is a sphere whose radius is
the size of a cross-linker (i.e., $d$, the interaction distance). With the definition
Eq.~{(\ref{eq:bracket_def})}, the operation $\lfloor\cdots\rceil^{ij}_{\Omega(\mathbf
x_i+s_i\mathbf p_i)}$ integrates
its argument over all geometrically possible crosslink interactions,
between filaments $i$ and $j$; see Fig.~{\ref{fig:cartoons}} (b).    
By Taylor expanding and keeping terms up to second order in the filament arc
length $(s_i, s_j)$, we find 
\begin{eqnarray}
    &&\mathbf F_{i}^\mathrm{(tot)} = 
    \sum_j
    \left\lfloor\left\{
        \begin{array}{c}
        1\\ 
        + (s_i \mathbf p_i- s_j \mathbf p_j) \cdot \nabla 
            \\
            +\frac{1}{2}(s_i^2\mathbf p_i \mathbf p_i + s_j^2\mathbf p_j \mathbf p_j) : \nabla \nabla 
    \end{array}
\right\}\delta(\mathbf x - \mathbf x_j) \mathbf
f_{ij}\right\rceil_{\Omega(\mathbf x_i)}^{ij}
    + \mathbf F_i^\mathrm{(drag)} \nonumber\\
    \label{eq:rod_force}
\end{eqnarray}
and the network stress
\begin{eqnarray}
    &&\mathbf \Sigma =  
    -\frac{1}2 \sum_{i,j}
    \left\lfloor
    \begin{array}{c}
    \delta(\mathbf x-\mathbf x_i)\delta(\mathbf x^\prime-\mathbf x_j)
    \\ 
\left(\mathbf x_i -\mathbf x_j +s_i \mathbf p_i - s_j \mathbf p_j\right)\mathbf f_{ij}
\end{array}
\right\rceil_{\Omega(\mathbf x_i)}^{ij}
    \label{eq:rod_stress}
\end{eqnarray}
where we used that $\mathbf f_{ij}=-\mathbf f_{ji}$.

\subsubsection{Torques}
Similarly, the angular momentum flux that crosslinkers exert between filaments can be written
as
\begin{equation}
\mathbf t_{ij} = \bar{ \mathbf t}_{ij}\left(s_i, s_j \right) + s_i\mathbf
p_i\times \mathbf f_{ij},
    \label{eq:torque_density}
\end{equation}
which dimensionally is a torque per unit area. Thus
\begin{equation}
    \mathbf T_{ij} 
    = 
    \lfloor\delta(\mathbf x - \mathbf x_j - s_j \mathbf p_j)\mathbf
    t_{ij}\rceil_{\Omega(\mathbf x_i + s_i \mathbf p_i)}
\end{equation}
which leads to
\begin{eqnarray}
    \mathbf T_{i}^\mathrm{(tot)} 
    = 
\mathbf T_i^\mathrm{(drag)}+
    \sum_j
    \left\lfloor
    \begin{array}{c}
        \delta(\mathbf x - \mathbf x_j)\left(\bar{\mathbf t}_{ij}  +
        s_i\mathbf p_i\times\mathbf f_{ij}\right)\\
        + (s_i\mathbf p_i -s_j
    \mathbf p_j)\cdot\nabla\delta(\mathbf x - \mathbf x_j)\left(\bar {\mathbf t}_{ij}  +
        s_i\mathbf p_i\times\mathbf f_{ij}\right)\\
        + \frac{1}{2}(s^2_i\mathbf p_i\mathbf p_i +s^2_j
    \mathbf p_j\mathbf p_j):\nabla\nabla\delta(\mathbf x - \mathbf x_j) \bar {\mathbf t}_{ij}
    \end{array}
    \right\rceil_{\Omega(\mathbf x_i)}^{ij}
    \nonumber\\
    \label{eq:rod_torque}
\end{eqnarray}
In the following we will consider crosslinks for which $\bar{
\mathbf{t}}_{ij} = 0$, for simplicity. 

\section{Filament-Filament interactions by crosslinks and collisions}
\label{sse:interactions}
We next discuss how filaments in highly crosslinked networks exchange linear and
angular momentum. Two types of interactions are important here: interactions mediated
by crosslinking molecules, which can be simple static linkers or active
molecular motors, and
steric interactions. We start by discussing the
former.

\subsection{Crosslinking interactions}
To describe crosslinking interactions, we propose a phenomenological model for
the stress $\mathbf f_{ij}$ that crosslinkers exert between the attachment
positions $s_i$ and $s_j$ on filaments $i$ and $j$. 
\begin{eqnarray}
    \mathbf f_{ij} &=& K(s_i, s_j, t)\left( \mathbf x_i +s_i\mathbf p_i -
    \mathbf x_j -s_j\mathbf p_j\right) 
    \nonumber\\
    &+& \gamma (s_i, s_j, t)\left( \mathbf v_i +s_i\dot{\mathbf p}_i - \mathbf
    v_j -s_j\dot{\mathbf p}_j\right)
    \nonumber\\
    &+& \left[ \sigma(s_i, s_j, t) \mathbf p_i - \sigma(s_j, s_i, t) \mathbf p_j \right ],
    \label{eq:generic_force_density}
\end{eqnarray}
The first term in this
model, with coefficient $K$, is proportional to the displacement between between
the attachment points,  $\mathbf x_i+s_i\mathbf p_i - \mathbf x_j -s_j\mathbf
p_j$, and captures the effects of crosslink elasticity and motor slow-down under
force. The second term, with coefficient $\gamma$, is proportional to $\mathbf
v_i +s_i\dot{\mathbf p}_i - \mathbf v_j -s_j\dot{\mathbf p}_j$, and captures
friction-like effects arising from velocity differences between the attachment
points. The last terms are motor forces that act along filament orientations
$\mathbf{p}_i$ and $\mathbf{p}_j$, with their coefficients $\sigma$ having
dimensions of stress.  Additional forces proportional to the relative rotation
rate between filaments, $\dot { \mathbf p}_i - \dot{ \mathbf p}_j$, are allowed
by symmetry, but are neglected here for simplicity. 

In general, the coefficients $K$, $\gamma$, and $\sigma$ are tensors
that depend on time, the relative orientations between microtubule $i$ and $j$
and the attachment positions $s_i, s_j$ on both filaments. In this work, we take
them to be scalar and independent of the relative orientation, for simplicity. 
Generalizing the calculations that follow to
include the dependences of $K, \gamma$ and $\sigma$ on $\mathbf p_i$
and $\mathbf p_j$ is straightforward but laborious and will be discussed in a
subsequent publication.  We emphasize that
Eq.~{(\ref{eq:generic_force_density})} is a statement about the expected average
effect of crosslinks in a dense local environment and is not a description
of individual crosslinking events. 

Inserting Eq.{(\ref{eq:generic_force_density})} into
Eqs.~{(\ref{eq:rod_force}, \ref{eq:rod_stress}, \ref{eq:rod_torque})} we find
that the stresses and forces collectively generated by crosslinks depend on
$s_{ij}$-moments of the form 
\begin{equation}
    X_{nm}(\mathbf x) =  \lfloor X(s_i, s_j) s_i^n s_j^m\rceil_{\Omega(\mathbf x)}^{ij},
    \label{eq:moment_definition}
\end{equation}
where $X=K$, $\gamma$, or $\sigma$. We refer to these as
crosslink moments. In principle, given Eqs.{(\ref{eq:rod_force},
\ref{eq:rod_stress}, \ref{eq:generic_force_density})} only the moments $X_{00}, X_{01},
X_{10}, X_{11}, X_{20}, X_{02}, X_{21},$ and $X_{12}$, contribute to the
stresses and forces in the filament network. We further note that $X_{11}$,
$X_{21}$ and $X_{12}$ are $\mathcal{O}(L^4)$, and can thus be neglected
without breaking asymptotic consistency. Moreover, $X_{20}$ and $X_{02}$ can be
expressed in terms of lower order moments since $X_{20} = X_{02} + \mathcal
{O}(L^4) = (L^2/12) X_{00} + \mathcal {O}(L^4)$. 

Finally, by construction $K(s_i, s_j) = K(s_j, s_i)$ and $\gamma(s_i, s_j) =
\gamma(s_j, s_i)$, and thus $\gamma_{01}=\gamma_{10}\equiv\gamma_1$ and
$K_{01}=K_{10} \equiv K_1$.  To further simplify our notation, we introduce $X_0
=X_{00}$. Explicit expressions for the seven crosslinking moments that contribute
to the continuum theory are given in the Appendix \ref{app:coeffcients}.
In summary, in the long wave length limit all forces and stresses in
the network can be expressed in terms of just a few moments, $K_0, K_1,
\gamma_0, \gamma_1, \sigma_0, \sigma_{01}, \sigma_{10}$. How different
crosslinker behavior set these moments will be discussed in
Section\ref{sse:macro}.

\subsection{Sterically mediated interactions}
In addition to crosslinker mediated forces and torques, steric interactions
between filaments generate momentum and angular momentum transfer in
the system. We model steric interactions by a free energy 
$E = \int_{\mathcal V}e(\mathbf p_i, \cdots, \mathbf x_i,\cdots) d^3x$ which
depends on all particle positions and orientations. The steric force is 
\begin{equation}
    \bar{\mathbf F}_i = -\frac{\delta E}{\delta \mathbf x_i},
\end{equation}
and the torque acting on it is
\begin{equation}
    \bar{\mathbf T}_i = -\frac{\delta E}{\delta \mathbf p_i}.
\end{equation}
This approach is commonly used throughout soft matter physics
\cite{martin1972unified,chaikin2000principles}. Common
choices for the free energy density $e$ are the ones proposed by Maier and Saupe
\cite{doi1988theory}, or Landau and DeGennes \cite{de1993physics}.

\section{Continuum Theory for highly crosslinked active networks}
\label{sse:continuum}
In the previous sections we derived a generic expression for the stresses and
forces acting in a network of filaments interacting through local forces and
torques, and proposed a phenomenological model for crosslink-driven interactions
between filaments. We now
combine these two and obtain expressions for the stresses, force, and torques
acting in a highly crosslinked filament network, and from there
derive equations of motion for the material. We start by introducing the
coarse-grain fields in terms of which our theory is phrased.

\subsection{Continuous Fields}
The coarse grained fields of relevance are the number density,
\begin{equation}
    \rho = \sum_i \delta(\mathbf x-\mathbf x_i),
    \label{eq:Density definition}
\end{equation}
the velocity $\mathbf v = \left<\mathbf v_i \right>$,
the polarity 
$\mathbf P =\left<\mathbf p_i \right>$,
the nematic-order tensor
$\mathcal Q =\left<\mathbf p_i\mathbf p_i \right>$,
and the third and fourth order tensors
$\mathcal T = \left<\mathbf p_i\mathbf p_i\mathbf p_i\right>$, and
$\mathcal S = \left<\mathbf p_i\mathbf p_i\mathbf p_i\mathbf p_i\right>$.
Here the brackets $\left<\cdot\right>$ signify the averaging operation
\begin{equation}
    \rho\left<\phi_i\right> = \sum_i \delta(\mathbf x -\mathbf x_i)
    \phi_i,
    \label{eq:averaging_definition}
\end{equation}
where $\phi_i$ is a dummy variable.
Furthermore, we define the tensors
$\mathbf j = \left<\mathbf p_i\left(  \mathbf v_i - \mathbf v\right)\right>$,
$\mathcal J = \left<\mathbf p_i\mathbf p_i\left(  \mathbf v_i - \mathbf v\right)\right>$,
$\mathcal H = \left<\mathbf p_i\dot{\mathbf p}_i\right>$,
and the rotation rate
$\mathbf \omega =\left<\dot{\mathbf p}_i\right>$.

\subsection{Stresses}
The presence of crosslinkers generates stresses in the material which,
through Eq.~{(\ref{eq:rod_stress}), depends on the crosslinking force density Eq.~{(\ref{eq:generic_force_density})}.
Following the nomenclature from Eq.~{(\ref{eq:generic_force_density})},
we write the material stress as
\begin{equation}
    \mathbf \Sigma = \mathbf \Sigma^{(K)} + \mathbf \Sigma^{(\gamma)} 
    + \mathbf \Sigma^{(V)} + \mathbf{\bar\Sigma},  
\end{equation}
where 
\begin{equation}
    \mathbf\Sigma^{(K)} = -\rho^2K_0\left(\alpha  \mathcal{I}
    +\frac{L^2}{12}\mathcal{Q}\right),
    \label{eq:StressK}
\end{equation}
is the stress due to the crosslink elasticity,
\begin{equation}
    \mathbf\Sigma^{(\gamma)} = -\rho^2 \left(\eta \nabla \mathbf v 
    + \gamma_1 \mathbf j + \gamma_0\frac{L^2}{12}\mathcal H \right),
    \label{eq:StressGamma}
\end{equation}
is the viscous like stress generated by crosslinkers,
and
\begin{equation}
    \mathbf\Sigma^{(V)} = -\rho^2 \left(\alpha \sigma_0 \nabla \mathbf P 
    + \sigma_{10} \mathcal Q - \sigma_{01}\mathbf P \mathbf P \right)
    \label{eq:StressV}
\end{equation}
is the stress generated by motor stepping. 
Here, we defined the network viscosity $\eta = \alpha\gamma_0$ and $\alpha =
\frac{3R^2}{10}$.

Finally, the steric (or Ericksen) stress obeys the Gibbs Duhem Relation 
\begin{equation}
    \nabla\cdot\bar{\mathbf\Sigma} = \rho\nabla\mu  + (\nabla \mathcal
    E):\mathcal Q.
\end{equation}
where
$\mu = -\frac{\delta e}{\delta \rho}$ is the chemical potential, and 
$\mathcal E = -\frac{\delta e}{\delta \mathcal Q}$ is the steric distortion
field. An explicit definition of $ \bar{\mathbf\Sigma}$ and the derivation of the
Gibbs Duhem relation are given in Appendix~{(\ref{app:steric})}. Note that for
simplicity, we chose that the steric free energy density $e$ depends only on
nematic order and not on
polarity.

\subsection{Forces}
We now calculate the forces acting on
filament $i$.
The total force $\mathbf F_i$ on filament $i$ is given by 
\begin{equation}
    \mathbf F_i = \mathbf F_i^{(K)} + \mathbf F_i^{(\gamma)} + \mathbf
    F_i^{(V)} + \bar{\mathbf F}_i + \mathbf F_i^\mathrm{(drag)},
    \label{eq:ForceAll}
\end{equation}
where 
\begin{eqnarray}
    \mathbf F^{(K)}_i &=& 
    (\nabla\rho)\cdot \frac{L^2}{12}K_0(\mathbf p_i\mathbf p_i -\mathcal{Q})
    -\frac{1}{\rho}\nabla\cdot\mathbf\Sigma^{(K)},
    \label{eq:ForceK}
\end{eqnarray}
is the elasticity driven force 
\begin{eqnarray}
    \mathbf F_i^{(\gamma)} &=& \gamma_0 \rho (\mathbf v_i -\mathbf v) 
    + \gamma_1 \rho (\dot{\mathbf p}_i -\mathbf\omega)
    \nonumber\\
    &+& \gamma_1 \left( \nabla \rho \right)\cdot\left[ \mathbf p_i\left(
            \mathbf v_i -\mathbf v \right) -\mathbf j - \mathbf P\left(
        \mathbf v_i -\mathbf v\right) \right] 
        \nonumber\\
        &+& \frac{L^2}{12}\gamma_0 \left( \nabla \rho \right)\cdot\left[\mathbf p_i
    \dot{\mathbf p}_i - \mathcal H \right]
        \nonumber\\
    &+& \frac{L^2}{12}\gamma_0 \left( \nabla\nabla \rho \right):\left[
        \mathbf p_i {\mathbf p}_i \left(\mathbf v_i-\mathbf v\right) - \mathcal
    J+\mathcal Q \left(\mathbf v_i-\mathbf v\right) \right]
        \nonumber\\
        &-& \frac{1}{\rho}\nabla\cdot\mathbf\Sigma^{(\gamma)}.
\label{eq:ForceGamma}
\end{eqnarray}
is the viscous like force,
and 
\begin{eqnarray}
    \mathbf F_i^{(V)} &=& \rho \sigma_0(\mathbf p_i - \mathbf P) 
    \nonumber\\
    &+& (\nabla \rho) \cdot \left[ \sigma_{10}\left(\mathbf p_i \mathbf p_i -\mathcal Q  \right)
    - \sigma_{01}\left( \mathbf p_i \mathbf P +  \mathbf P \mathbf p_i - 2\mathbf P\mathbf P\right)\right ]
    \nonumber\\
    &+& \frac{L^2}{12} \sigma_0 (\nabla \nabla \rho):\left[ \mathbf p_i\mathbf p_i\mathbf p_i +
    \mathcal Q \mathbf p_i - \mathbf p_i \mathbf p_i \mathbf P -
\mathcal T \right]
\nonumber\\
&-&\frac{1}\rho \nabla\cdot \mathbf\Sigma^{(V)}.
\label{eq:ForceV}
\end{eqnarray}
is the motor force.
Finally, 
\begin{equation}
    \bar {\mathbf{F}}_i = -\frac{\nabla\mathcal E}{\rho}:(\mathbf
    p_i \mathbf p_i -\mathcal Q)   -\frac{1}\rho
\nabla\cdot\bar{\mathbf\Sigma}, 
\end{equation}
is the steric force on filament $i$,
where we again chose $e$ to only depend on nematic order and not on polarity.

\subsection{Crosslinker induced Torque}
We next calculate the torques acting on filament $i$. The total
torque acting on filament $i$ is 
\begin{equation}
    \mathbf T_i = \mathbf T_i^{(\gamma)} + \mathbf T_i^{(V)}
    +\bar{\mathbf T}_i+ \mathbf T_i^{(\mathrm{drag})}
        \label{eq:filament_total_torque}
\end{equation}
Note, that crosslinker elasticity does not contribute.
Here
    \begin{eqnarray}
        \mathbf T_i^{(\gamma)} &=&         
        \gamma_1 \rho\mathbf p_i\times\left( \mathbf v_i -\mathbf v \right) 
    + \frac{L^2}{12}\gamma_0 \rho\mathbf p_i\times\dot{\mathbf p}_i
            \nonumber\\
            &+& \frac{L^2}{12}\gamma_0 \mathbf p_i\times\left(\mathbf p_i\cdot \nabla \rho
            \right)\left( \mathbf v_i - \mathbf v \right)  \nonumber\\
        &-&\frac{L^2}{12}\gamma_0\rho\mathbf p_i\times(\mathbf p_i\cdot\nabla\mathbf v)
    \label{eq:TorqueGamma}
\end{eqnarray}
and
\begin{eqnarray}
    \mathbf T^{(V)}_i &=& -\rho\mathbf p_i\times\left(\sigma_{01}\mathbf P +
        \frac{L^2}{12}\sigma_0\mathbf
    p_i\cdot\nabla\mathbf P\right)
-\frac{L^2}{12}\sigma_0\mathbf p_i\times (\mathbf p_i\cdot\nabla\rho)\mathbf P
\end{eqnarray}
are the viscous and motor torques, respectively.
Steric interactions contribute to the torque
\begin{equation}
    \bar{\mathbf{T}}_i = \mathbf p_i\times \frac{\mathcal E}{\rho}\cdot\mathbf p_i.
\end{equation}

\subsection{Equations of Motion}
To find equations of motion for the highly crosslinked network, we use
Eqs.~{(\ref{eq:ForceAll}, \ref{eq:ForceK}, \ref{eq:ForceGamma},
\ref{eq:ForceV})}, and obtain 
\begin{eqnarray}
    \mathbf v_i-\mathbf v &=&
    - \frac{\sigma_0}{\gamma_0} (\mathbf p_i-\mathbf P)
    - \frac{1}{\rho\gamma_0}\left(\mathbf F^\mathrm{(drag)}_i - f/\rho\right)
    + \mathcal{O}\left( L^2 \right),
    \label{eq:low_order_vi}
\end{eqnarray}
which will be a useful low-order approximation to $\mathbf v_i -\mathbf v$.
Note too that we have dropped steric forces, since $\nabla \mathcal E
/\rho$ scales with the inverse of the system size, which is much larger than $L$.
Using Eq.~{(\ref{eq:low_order_vi})} in Eq.~{(\ref{eq:filament_total_torque})}
we find the equation of motion for filament rotations,
\begin{eqnarray}
    \dot{\mathbf p}_i &=& \left( \mathcal{I} -\mathbf p_i \mathbf p_i
    \right)\cdot
    \left\{
        \begin{array}{c}
        \mathbf p_i\cdot\mathcal U 
        \\+ \frac{12}{\gamma_0L^2\rho^2}\mathbf p_i\cdot \mathcal E
        \\+
    \frac{12 }{\gamma_0L^2}A^{(\mathbf P)} \mathbf P 
\end{array}
\right\},
    \label{eq:kinetic_angular}
\end{eqnarray}
where we neglect drag mediated terms, which are subdominant at high density,
for simplicity. A detailed calculation, and expressions which includes drag terms,
is given in Appendix~\ref{app:calculate}.
Here, 
\begin{equation}
    \mathcal U = \nabla\mathbf v +\frac{\sigma_0}{\gamma_0}\nabla\mathbf P,
\end{equation}
is the active strain rate tensor, which consists of the consists of the strain
rate and vorticity $\nabla\mathbf v$ and an active polar contribution
$\nabla\mathbf P$. 
Moreover 
\begin{equation}
    A^{(\mathbf P)} = \sigma_{01}-\sigma_0\frac{\gamma_1}{\gamma_0}.
    \label{eq:App}
\end{equation}
is the polar activity coefficient.
The filament velocities are given by
\begin{eqnarray}
    \mathbf v_i -\mathbf v &=& -\frac{\sigma_0}{\gamma_0}\left( \mathbf p_i -\mathbf
    P\right) 
    \nonumber\\
    &-&\frac{\gamma_1}{\gamma_0}\left((\mathbf p_i-\mathbf P)\cdot\mathcal{U}  
    -(\mathbf p_i\mathbf p_i\mathbf p_i-\mathcal T):\mathcal{U}\right)
    \nonumber\\
    &-&\frac{12 \gamma_1}{ L^2 \rho^2\gamma_0^2}\left((\mathbf p_i-\mathbf
        P)\cdot\mathcal{E}  
    -(\mathbf p_i\mathbf p_i\mathbf p_i-\mathcal T):\mathcal{E}\right)
    \nonumber\\
    &+&\frac{12 \gamma_1}{L^2\gamma_0^2}A^{(\mathbf P)}\left( \mathbf p_i\mathbf
    p_i-\mathcal Q \right)\cdot \mathbf P,
    \label{eq:v_final}
\end{eqnarray}
where we used Eqs.~{(\ref{eq:low_order_vi}, \ref{eq:kinetic_angular})} in Eq.~{(\ref{eq:ForceAll})}.
In Eq.~{(\ref{eq:v_final})}, we ignored terms proportional to density gradients,
for simplicity. The full expression is given in Appendix~\ref{app:calculate}.
After some further algebra (see Appendix~\ref{app:calculate}),
we arrive at an expression for the material stress in terms of the current
distribution of filaments,
\begin{eqnarray}
    \mathbf\Sigma &=&
    -\rho^2\left(\chi:\mathcal U +\alpha K_0\mathcal{I}
       + A^{(\mathcal Q)}\mathcal Q
    -A^{(\mathbf P)}\mathcal
T\cdot \mathbf P \right) 
    + \mathbf\Sigma^\mathrm{(S)},
    \label{eq:stress_final}
\end{eqnarray}
where 
\begin{equation}
    \chi_{\alpha\beta\gamma\mu} = \eta\delta_{\alpha\gamma}\delta_{\beta\mu}
    +
    \frac{L^2}{12}\gamma_0\left(\mathcal Q_{\alpha\gamma}\delta_{\beta\mu} \mathcal
    - S_{\alpha\beta\gamma\mu}\right),
    \label{eq:generalized_viscosity}
\end{equation}
is the anisotropic viscosity tensor,
\begin{equation}
    A^{(\mathcal Q)} = \sigma_{10}-\sigma_0\frac{\gamma_1}{\gamma_0} +\frac{L^2}{12}K_0 
    \label{eq:AQ}
\end{equation}
is the nematic activity coefficient,
and 
\begin{equation}
    \mathbf\Sigma^\mathrm{(S)}_{\alpha\beta} = \bar{\mathbf\Sigma}_{\alpha\beta}
    -\left( \mathcal Q_{\alpha\gamma}\delta_{\beta\mu} \mathcal
    - S_{\alpha\beta\gamma\mu} \right)\mathcal E_{\gamma\mu}.
    \label{eq:steric_stress_final}
\end{equation}
is the steric stress tensor.
Together Eqs.~{(\ref{eq:smoluchowski}, \ref{eq:kinetic_angular},
 \ref{eq:v_final}, \ref{eq:stress_final})} define a full kinetic theory for the
highly crosslinked active network.

\section{Designing materials by choosing crosslinks}
\label{sse:macro}
Eqs.~{(\ref{eq:smoluchowski}, \ref{eq:kinetic_angular},
\ref{eq:stress_final}, \ref{eq:v_final})} define a full kinetic theory for 
highly crosslinked active networks. This theory has the same active stresses
known from symmetry based theories for active
materials\cite{marchetti2013hydrodynamics, kruse2005generic,
furthauer2012active} and thus can give rise to the same rich phenomenology. 
Since our framework derives these stresses from microscale properties of the
constituents of the material it enables us to make predictions on how the
microscopic properties of the network constituents affect its large scale
behavior.  We first discuss how motor properties set
crosslink moments in Eq.~{(\ref{eq:generic_force_density})}. We
then study how these crosslink properties impact the large scale properties of
the material.

\subsection{Tuning Crosslink-Moments}
\begin{figure}[h]
    \centering
    \includegraphics[width=0.75\columnwidth]{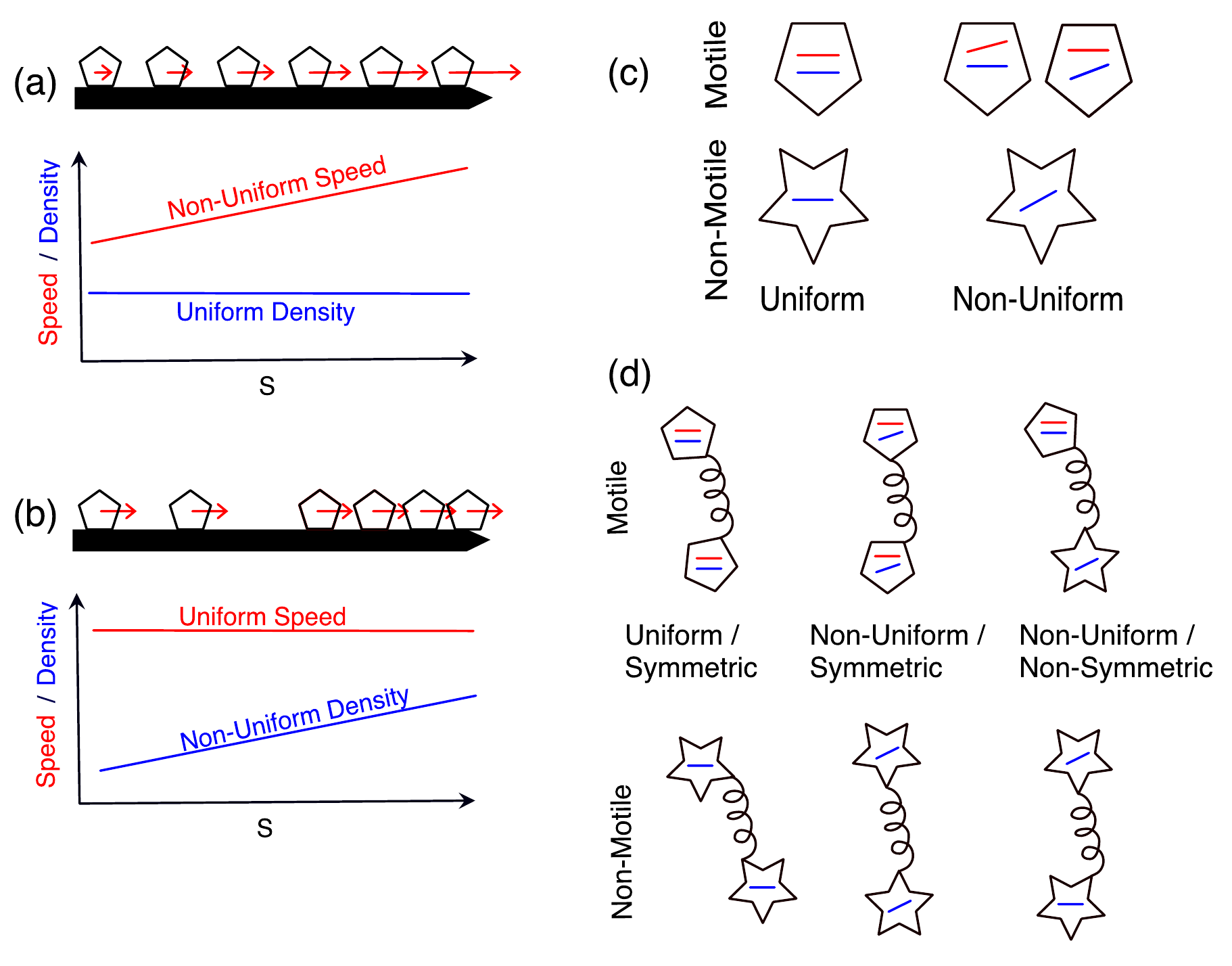}
    \caption{(a, b) Populations of crosslink heads are characterized by the density with which they
    bind a filament along its arclength $s$ and the speed at which they move
    when force free. Two different head types, one with non-uniform speed but
    uniform density (a) another with uniform speed and non-uniform density (b)
    are shown. In (c) we list some possible crosslink heads. Red and Blue
lines illustrate the change of crosslink speed and density with $s$,
respectively. In (d) we illustrate example crosslinks which consist of two heads
and a linker.}
\label{fig:motor_types}
\end{figure}

The coefficients in Eq.~{(\ref{eq:generic_force_density})} arise from a
distribution of active and passive crosslinks that act between filaments. Consider an
ensemble of crosslinking molecules, each consisting of two heads $a$ and
$b$, joined by a spring-like linker; see
Fig.~{(\ref{fig:motor_types})}. For any small volume in an active network, we
can count the number densities $\xi_a(s)$, and $\xi_b(s)$ of $a$ and $b$ heads of
doubly-bound crosslinks that are attached to a filament at arc-length position
$s$. In an idealized experiment $\xi_a(s)$ and $\xi_b(s)$ could be
determined by recording the positions of motor heads on filaments.
The number-density $\xi_{ab}(s_i, s_j)$ of $a$ heads at position $s_i$ on
microtubule $i$ connected to $b$ heads at position $s_j$ on
microtubule $j$ is then given by
\begin{eqnarray}
    \xi_{ab}(s_i, s_j) = 
    \frac{\xi_a(s_i)\xi_b(s_j)}
    {N_b^{(i)}(s_i)}
    \label{eq:binding_xiab}
\end{eqnarray}
where $N^{(i)}_b(s_i)$ counts the $b$ heads that an $a$-head attached at
position $s_i$ on filament $i$ could be connected to given the crosslink size.
It obeys
\begin{equation}
    N^{(i)}_b(s_i) =    
    \sum\limits_{k\ne i}
    \int\limits_{-L/2}^{L/2} ds_k
    \int\limits_{\Omega(\mathbf x_i + s_i\mathbf
    p_i)}d\mathbf x^3 \xi_b(s_k)\delta(\mathbf x_k + s_k \mathbf p_k -\mathbf x).
\end{equation}
Analogous definitions for $\xi_{ba}(s_i, s_j)$ and $N_a^{(i)}(s_i)$ are implied.
It follows naturally that $\xi(s_i, s_j) =\xi_{ab}(s_i, s_j) + \xi_{ba}(s_i,
s_j)$ is the total number density of crosslinks acting between filaments $i$ and
$j$ at the arclength positions $s_i$, $s_j$.

Now let $V_a(s), V_b(s)$ be the load-free velocities of motor-heads $a, b$
moving along filaments.  Here, $V_a(s), V_b(s)$ are functions of the arc-length
position $s$. Like $\xi_a$ and $\xi_b$, they are in principle measurable. 
With these definitions, the force per unit surface that attached motors
exert is
\begin{eqnarray}
    &&\mathbf f_{ij} = 
    -\Gamma \xi(s_i, s_j)
    \left(\mathbf v_i +s_i\dot{\mathbf p}_i - \mathbf v_j +s_j\dot{\mathbf p}_j \right)
    \nonumber\\
    &&-\kappa \xi(s_i, s_j)\left(\mathbf x_i +s_i\mathbf p_i - \mathbf x_j +s_j\mathbf p_j \right)
    \nonumber\\
    &&-\Gamma \left(\left[\xi_{ab}(s_i, s_j)V_a(s_i) + \xi_{ba}(s_i, s_j)V_b(s_i)\right]\mathbf p_i  \right)
    \nonumber\\
    &&+\Gamma \left(\left[\xi_{ab}(s_j, s_i)V_a(s_j) +
    \xi_{ba}(s_j,s_i)V_b(s_j)\right]\mathbf p_j  \right), 
\end{eqnarray}
where $\Gamma$ is an effective linear friction
coefficient between the two attachment points and
$\kappa$ is an effective spring constant. They depend on the microscopic
properties of motors, filaments, and the concentrations of both and their
regulators. 
In general, $\Gamma$ and $\kappa$ are second rank tensors,
which depend on the relative orientations of filaments. Here we take them to be
scalar, for simplicity and consistency with earlier assumptions.
By comparing to Eq.~{(\ref{eq:generic_force_density})} we identify
\begin{equation}
    \gamma(s_i, s_j) = -\Gamma \xi(s_i,s_j),
    \label{eq:gamma_def}
\end{equation}
\begin{equation}
    K(s_i, s_j) = -\kappa \xi(s_i,s_j),
    \label{eq:K_def}
\end{equation}
and
\begin{eqnarray}
    \sigma(s_i, s_j) &=& -\Gamma \xi_{ab}(s_i, s_j)V_a(s_i)  
    +\Gamma \xi_{ba}(s_i,s_j)V_b(s_i).
    \label{eq:sigma_def}
\end{eqnarray}

\begin{table}
\begin{centering}
\hskip-1.25cm
\begin{tabular}{|c|c|c|c|c|c|}
    \hline
     &$\gamma_0$, $K_0$ & $\gamma_1$, $K_1$ & $\sigma_0$& $\sigma_{10}$& 
    $\sigma_{01}$ \\ 
    \hline
    \includegraphics[scale=.75]{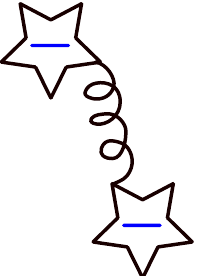}
    \begin{tabular}{c} symmetric \\ uniform \\  non-motile \end{tabular} 
    & yes & no & no & no & no \\
    \hline
    \includegraphics[scale=.75]{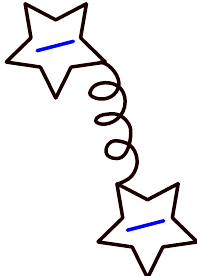}
    \begin{tabular}{c} non-symmetric \\ uniform \\  non-motile \end{tabular} 
    & yes & yes & no & no & no \\
    \hline
    \includegraphics[scale=.75]{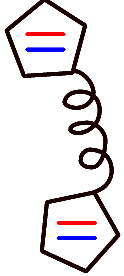} 
    \begin{tabular}{c} symmetric \\ uniform \\  motor \end{tabular} 
    & yes & no & yes & no & no \\
    \hline
    \includegraphics[scale=.75]{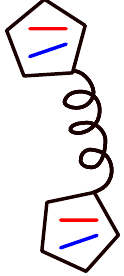}
    \begin{tabular}{c} symmetric \\ non-uniform \\  motor \end{tabular} 
    & yes & yes & yes &
    \begin{tabular}{c} yes \\ $\sigma_{10}=\sigma_1$ \end{tabular} & 
    \begin{tabular}{c} yes \\ $\sigma_{01}=\sigma_1$ \end{tabular}\\
    \hline
    \includegraphics[scale=.75]{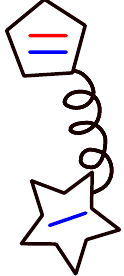}
    \begin{tabular}{c} non-symmetric \\ non-uniform \\  motor \end{tabular} 
    & yes & yes & yes & yes
     & yes \\
    \hline
\end{tabular}
\end{centering}
\caption{Table summarizing which crosslink moments different crosslink types
generate.}\label{tab:motor_to_moments}
\end{table}

Using Eqs.~{(\ref{eq:gamma_def}, \ref{eq:K_def},
\ref{eq:sigma_def})}, we now discuss some important classes of crosslinking
molecules. We consider crosslinks whose heads can be motile or non-motile, the
binding and walking properties can act uniformly or non-uniformly along
filaments, and the two heads of the crosslink can be the same (symmetric
crosslink) or different (non-symmetric crosslink). Figure~{\ref{fig:motor_types}} 
maps how varying crosslink types can be
constructed, while Table \ref{tab:motor_to_moments} lists the moments to which
different classes of crosslinks contribute. 

{\it Non-motile crosslinks} are crosslinks that do not actively move, i.e. $V_a
= V_b = 0$. Examples of non-motile crosslinks in cytoskeletal systems are the
actin bundlers such as fascin, or microtubule crosslinks such as Ase1p
\cite{alberts_2002_book}. 
While these types of crosslinks are not necessarily passive, since
the way they binding or unbind can break detailed balance, that their
attached heads do not walk along filaments implies that
$\sigma_0 = \sigma_{10} = \sigma_{01} = 0$. Non-motile crosslinks change the
material properties of the material by contributing to the crosslink moments
$\gamma_0, \gamma_1$ and $K_0, K_1$. Some non-motile crosslinks
bind non-specifically along filaments they interact with, giving uniform distributions. 
For these $\gamma_1 = K_1 = 0$. 
Others preferentially associate to filament ends, and thus bind non-uniformly.
For these $\gamma_1$ and $K_1$ are positive. Note that the two heads of a
non-motile crosslink can be identical (symmetric) or not (non-symmetric).
Given the symmetric structure of Eqs~{(\ref{eq:gamma_def}, \ref{eq:K_def})}
mechanically a non-symmetric non-motile crosslink behaves the same as a
symmetric non-motile crosslink. 
 and 
{\it Symmetric Motor crosslinks} are motor molecules whose two heads 
have identical properties, i.e. $V_a = V_b = V$ and $\xi_a = \xi_b = \xi$. 
Examples are the
microtubule motor molecule Eg-5 kinesin, and the Kinesin-2 motor construct
popularized by many in-vitro experiments\cite{sanchez2012spontaneous}.
Symmetric motors contribute to the large-scale properties of the material by
generating motor forces. In particular they contribute to the
crosslink moments $\sigma_0$, $\sigma_{10}$, and $\sigma_{01}$.  From
Eq.~{(\ref{eq:sigma_def})} it is easy to see that $\sigma_0 = V_0 \gamma_0 +V_1
\gamma_1/L^2$, where we defined the moments of the motor velocity $V(s_i, s_j)$
using Eq.~{(\ref{eq:moment_definition})}.
Some symmetric motor proteins preferentially associate to filament
ends, and display end-clustering behavior, where their walking speed depends 
on the position at which they are attached to filaments. Motors that do either
of these also generate a contribution to $\sigma_{10}$ and $\sigma_{01}$.  
Since both motor heads are identical we have $\sigma_{10} = \sigma_{01} \equiv \sigma_1$
and from Eq.~{(\ref{eq:sigma_def}) we find that $\sigma_1 = \gamma_1 V_0 +V_1
    \gamma_0$.

    {\it Non-Symmetric motor crosslinks} are motor molecules whose two heads have
differing properties. An example is the microtubule-associated motor dynein,
that consists of a non-motile end that clusters near microtubule minus-ends and a
walking head that binds to nearby microtubules whenever they are within
reach \cite{foster2015active, foster2017connecting}. A consequence of motors
being non-symmetric is that $\sigma_{10} \ne \sigma_{01}$. Since non-symmetric
motors can break the symmetry between the two heads in a variety of ways we
spell out the consequences for a few cases.
Let us first consider a crosslinker with one head $a$ that acts as a
passive crosslink ($V_a = 0$) and a second head $b$ that acts as a motor, moving
with the stepping speed $V_b=V$. For such a crosslink $\sigma_0 = \gamma_0
V_0/2$. If both heads are distributed uniformly along filaments and their $V$ is
position independent then $\sigma_{01} = \sigma_{10} = 0$. If the walking
$b$-head is distributed nonuniformly ($\xi_b =\xi_b(s), \xi_a=$ constant)
then $\sigma_{10} = \gamma_1 V_0$  and $\sigma_{01} = 0$.  Conversely, if the
static $a$-head has a patterned distribution ($\xi_a =\xi_a(s), \xi_b=$
constant) then $\sigma_{01} = \gamma_1 V_0, \sigma_{10} = 0$.  Finally, we note
that if both heads are distributed uniformly along the filament ($\xi_a =\xi_b =
$constant), but the walking $b$-head of the motor changes its speed as function
of position then $\sigma_{10} = V_1 \gamma_0/2$ and $\sigma_{01} = 0$.

Note that stresses and forces are additive. Thus it may be possible
to design specific crosslink moments by designing mixtures of different
crosslinkers. For instance mixing a non-motile crosslink that has specific
binding to a filament solution might allow to change just $\gamma_0$ and
$\gamma_1$ in a targeted way. We will elaborate on some of these possibilities
in what follows. 

\subsection{Tuning viscosity}
We now discuss how microscopic processes shape the overall magnitude of the
viscosity tensor $\chi$. From Eq.~{(\ref{eq:generalized_viscosity})} and
remembering that $\eta = 3R^2/10 \gamma_0$, it is apparent that the overall
viscosity of the material is proportional to the number of crosslinking
interactions and their resistance to the relative motion of filaments, 
quantified by the friction coefficient $\rho^2\gamma_0$. Furthermore, $\gamma_0$ itself
scales as the squared filament length $L$, and the cubed crosslink size
$R$ (see the definition in Appendix~\ref{app:coeffcients}), which, with $\rho^2$, sets 
the overall scale of the viscosity as $\rho^2L^2R^3$.

We next show how micro-scale properties of network constituents shape 
the anisotropy of $\chi$ ; see Eq.~{(\ref{eq:generalized_viscosity}). To
characterize this we define the anisotropy ratio $a$ as 
\begin{equation}
    a = \frac{L^2\gamma_0}{12\eta} = \frac{5}{18}\frac{L^2}{R^2},
\end{equation}
which is the ratio of the magnitudes of the isotropic part of
$\chi_{\alpha\beta\gamma\mu}$, that is
$\eta \delta_{\alpha\gamma}\delta_{\beta\mu}$, and its anisotropic part $\gamma_0
 L^2/12(\mathcal
Q_{\alpha\gamma}\delta_{\beta\mu}-\mathcal{S}_{\alpha\beta\gamma\mu})$.
Most apparently the anisotropy ratio will be large if the typical filament length $L$ is
large compared to the motor interaction range $R$. This is typically the case in
microtubule based systems, as microtubules are often microns long and
interact via motor groups that are a few tens of nano-meters in scale
\cite{alberts_2002_book}.  
Conversely, in actomyosin systems filaments are often
shorter (hundreds of nano-meters) and motors-clusters called
mini-filaments, can have sizes similar to the filament lengths
\cite{alberts_2002_book}.  
The anisotropy of the viscous stress is not exclusive to active systems and has
been described before in the context of similar passive systems, such as liquid
crystals and liquid crystal polymers
\cite{de1993physics,chaikin2000principles,doi1988theory}.

\subsection{Tuning the active self-strain}

\begin{table}
\hskip-1.3cm
\begin{centering}
\begin{tabular}{|c|c|c|}
    \hline
    Mixture
    & Active Strain, $\sigma_0/\gamma_0$ & \begin{tabular}{c} Active Pressure,
    $\Pi^\mathrm{(A)}$ \\ Axial stress, $\bar{S}$ \end{tabular} \\ 
    \hline
    \includegraphics[scale=.5]{msu.pdf}& $\frac{\sigma_0}{\gamma_0}=V_0$
    & no \\
    \hline
    \includegraphics[scale=.5]{msnu.pdf}&
    $\frac{\sigma_0}{\gamma_0}=V_0$ & no \\
    \hline
    \includegraphics[scale=.5]{mnsnu.pdf}&
    $\frac{\sigma_0}{\gamma_0}=\frac{V^a_0+V^b_0}{2}$ &
    \includegraphics[scale=.4]{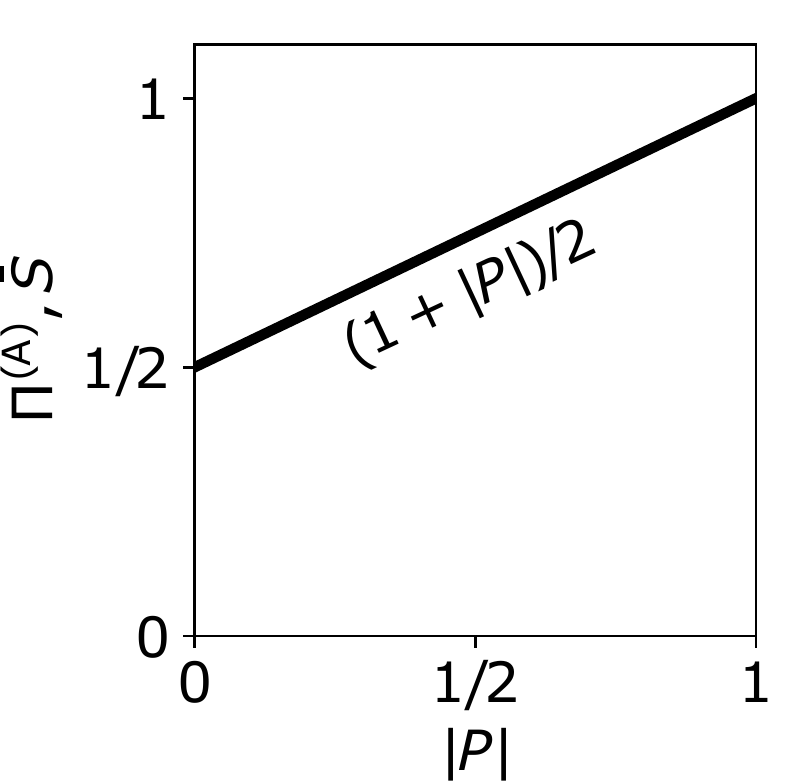}
    \\
    \hline
    \includegraphics[scale=.5]{msu.pdf}
    +
    \includegraphics[scale=.5]{csu.pdf} & 
    \quad
    \includegraphics[scale=.4]{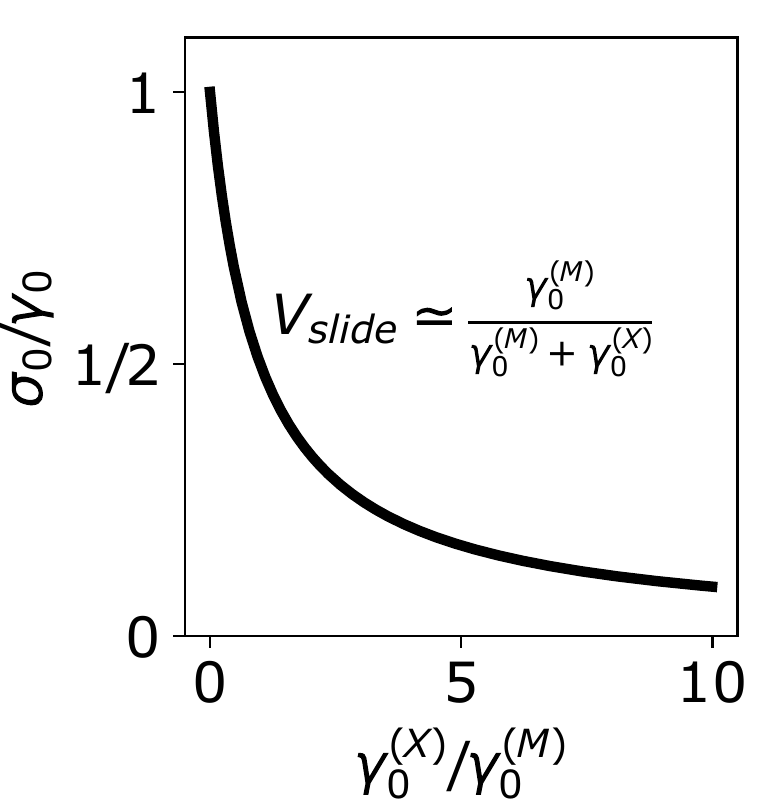} 
    \quad
    & no  \\
    \hline
    \begin{tabular}{c}
    \includegraphics[scale=.5]{msu.pdf}
    +
    \includegraphics[scale=.5]{csnu.pdf}
    \\ or \\
    \includegraphics[scale=.5]{msnu.pdf}
    +
    \includegraphics[scale=.5]{csu.pdf} 
    \end{tabular}
    &
    \includegraphics[scale=.4]{slgr.pdf} 
    & 
    \quad
    \includegraphics[scale=.4]{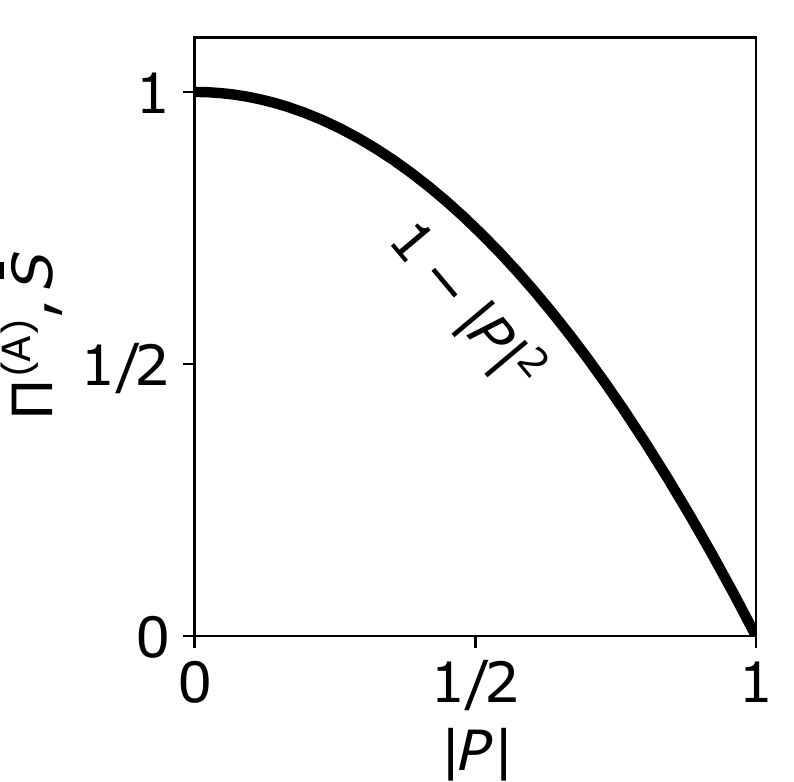}
    \quad
    \\
    \hline
\end{tabular}
\end{centering}
\caption{Active pressure and strain generated by different crosslink types
    and mixtures. In the plots pertaining to the active strain rate $\gamma_0
    =\gamma_0^{(M)}+\gamma_0^{(X)}$ where $\gamma_0^{(M)}$ denotes the is the
    part of $\gamma_0$ induced by mobile and $\gamma_0^{(X)}$ denotes
    contribution from non-mobile crosslinkers. The filament sliding velocity
    expected in a stress free system is $V_{slide} = \sigma_0/\gamma_0$ and is
    given units of the force free speed of immobile crosslinks and describe the
    expected speed of filament sliding in the material.  Moreover, $\bar S=
    |\Pi^\mathrm{(A)}/q|$ is the magnitude of the motor-stepping induced axial
    stress, i.e of the axial stress in the limit $K_0\to0$.}
\label{tab:motor_to_effects}
\end{table}

The viscous stress in highly
crosslinked networks is given by $\chi:\mathcal U$, where $\mathcal U = \nabla
\mathbf v + (\sigma_0/\gamma_0)\nabla\mathbf P$ takes the role of the strain-rate
in passive materials, but with an active contribution
$(\sigma_0/\gamma_0)\nabla\mathbf P$. Thus, internally driven materials can
exhibit active self-straining. 

In particular a
material in which each filament moves with the velocity $\mathbf v_i =
-\sigma_0/\gamma_0\mathbf p_i + \mathbf C$, where $\mathbf C$ is a constant
vector that is sets the net speed of the material in the frame of reference, has
$\mathcal U =0$, and thus zero viscous stress. In such a material filaments can
slide past each other at a speed
$\sigma_0/\gamma_0$ without stressing the material. Notably, the sliding speed
is independent of the local polarity and nematic order of the
material \cite{furthauer2019self}.

The crosslink moments that contribute to the active straining behavior are
$\sigma_0$ and $\gamma_0$. In active filament networks with a single type of
crosslink $\sigma_0/\gamma_0\simeq V_0$, regardless of crosslink
concentration.  Thus for single-crosslinker systems, the magnitude of
self-straining is independent of the motor concentration \cite{furthauer2019self}. 

Self-straining can be tuned in mixtures of crosslinks. For instance the
addition of a non-motile crosslinker can increase $\gamma_0$, while leaving
$\sigma_0$ unchanged. In this way self-straining can be relatively suppressed. 
In table \ref{tab:motor_to_effects} we plot the expected active strain-rate for
materials actuated by mixtures of immotile and motor crosslinks. In such a
material $\gamma_0 = \gamma_0^{(M)}+\gamma_0^{(X)}$ where $\gamma_0^{(M)}$ denotes the 
part of $\gamma_0$ induced by motile crosslinkers and $\gamma_0^{(X)}$ denotes
that from non-motile crosslinkers. The resulting velocity $V_{slide}$
with which a filament slides through the material will scale as
$V_{slide}\simeq\gamma_0^{(M)}/( \gamma_0^{(M)}+\gamma_0^{(X)})$; see
Table~\ref{tab:motor_to_effects}.

\subsection{Tuning the Active Pressure}
Many active networks spontaneously contract \cite{foster2015active} or
expand \cite{sanchez2012spontaneous}. 
We now study the motor properties that
enable these behaviors.

An active material with stress free boundary conditions, can spontaneously
contract if its self-pressure, \begin{equation}
\Pi = \mathrm{Tr}\left( \mathbf\Sigma + \rho^2\chi:\mathcal U \right). 
\end{equation}
is negative.
Conversely the material can spontaneously extend if $\Pi$ is positive. 
We can also write
\begin{equation}
    \Pi = \Pi^\mathrm{(A)} + \Pi^\mathrm{(S)}
\end{equation}
where $\Pi^\mathrm{(S)} = \mathrm{Tr}\left(\mathbf \Sigma^\mathrm{(S)} \right)$
is the sterically mediated pressure, and $\Pi^\mathrm{(A)}$ is the 
activity driven pressure (or active pressure) given by
\begin{equation}
    \Pi^\mathrm{(A)} = -\rho^2\left(\alpha K_0 + A^{(\mathcal Q)} - A^{(\mathbf
    P)} |\mathbf P|^2 \right).
\label{eq:mpress}
\end{equation}
see Eq.~{(\ref{eq:stress_final})}.
Here and in the following we approximated $\mathrm{Tr}(\mathcal T\cdot\mathbf P)\simeq
|\mathbf P|^2$ for simplicity.  We ask which properties of crosslinks
set the active pressure and how its sign can be chosen.

We first discuss how interaction elasticity impacts the active pressure
$\Pi^{(A)}$ in
the absence of motile crosslinks, i.e. when $\sigma_0 = \sigma_{10} = \sigma_{01} =
0$.  In this case, Eq.~{(\ref{eq:mpress})} simplifies to  $\Pi^\mathrm{(A)} =
-\rho^2(\alpha + L^2/12)K_0$, where we used Eq.~{(\ref{eq:AQ})}. Thus, even in the
absence of motile crosslinks, active pressure can be generated. This can be
tuned by changing the effective spring constant $K_0$. We note that
$\Pi^\mathrm{(A)}+\Pi^\mathrm{(S)}=0$ when crosslink binding-unbinding obeys
detailed balance and the system is in equilibrium. The moment $K_0$ can have
either sign when detailed balance is broken. Microscopically this effect could
be achieved, for instance, by a crosslinker in which active processes change the
rest length of a spring-like linker between the two heads once they bind to
filaments.

We next discuss the contributions of motor motility to the active pressure.
To start, we study a simplified apolar (i.e. $\mathbf P=0$) system where $K=0$.
In such a system the active pressure is given by
\begin{equation}
    \Pi^\mathrm{(A)} = -\rho^2\left(\sigma_{10} -
\sigma_0\frac{\gamma_1}{\gamma_0} \right)
\label{eq:pi_motors_only_nema}
\end{equation}
We ask how motor properties set the value and sign of this parameter
combination. 

We first point out that generating  active pressure by motor stepping requires
that either $\sigma_{10}$ or $\gamma_1$ are non-zero. This means that generating
active pressure requires breaking the uniformity of binding or walking
properties along the filament. A crosslink which has two heads that act
uniformly can thus not generate
active pressure on its own. However, when operating in conjunction with a
passive crosslink that preferentially binds either end of the filament, the same
motor can generate an active pressure. This pressure will be contractile if the
non-motile crosslinks couple the end that the motor walks towards ($\gamma_1$
and $\sigma_0$ have the same sign) and extensile if they couple the other
($\gamma_1$ and $\sigma_0$ have opposite signs). In summary, a motor crosslink
that acts the same everywhere along the filaments it couples does not generate
active pressure on its own. However, it can do so when mixed with a passive
crosslink that acts non-uniformly.

We next ask if a system with just one type of non-uniformly acting crosslink can
generate active pressure.  To start, consider {\it symmetric motor crosslinks},
i.e a motor consisting of two heads with identical (but non-uniform) properties. 
We then have $\sigma_{01} =
\sigma_{10} = \gamma_1 V_0 + \gamma_0 V_1$ and $\sigma_0 = V_0 \gamma_0 + V_1
\gamma_1 /L^2$. Using this in Eq.~{(\ref{eq:pi_motors_only_nema})} and dropping
the term proportional to $\gamma_1^2$ (higher order in this case) we find that
such symmetric motor crosslinks generate no contribution
to the active pressure when operating alone.  However when operating in concert
with a non-motile crosslink, even one that binds filaments uniformly,
they can generate and active pressure. The sign of the active
pressure is set by the particular asymmetry of motor binding and motion. The system is
contractile if motors cluster or speed up near the end towards which they walk, and extensile
if they cluster or accelerate near the end that they walk from. 
Our prediction that many motor molecules can only generate active pressure in
the presence of an additional crosslink, might explain observation on
acto-myosin gels, which have been shown to contract only when
combined a passive crosslink operate in concert with the motor myosin
\cite{ennomani2016architecture}. 

We next ask if {\it non-symmetric motor crosslinks} can generate active
pressure. Consider a crosslink with one immobile and one walking head. For such
a crosslink $\sigma_0 = \gamma_0 V_0/2$. If the immobile head preferentially
binds near one filament end, while the walking head attaches everywhere
uniformly, then $\sigma_{10} = \gamma_1 V_0$ and $\sigma_{01} = 0$. For such a motor we
predict an active pressure proportional to $V_0/2$. The active pressure will be
contractile if the static ends bind near the end that the motor head walks to
and extensile if the situation is reversed. The motor dynein has been
suggested to consist of an immobile head that attaches near microtubule minus ends
and a walking head that grabs other microtubules and walks towards their minus
ends. Our theory suggests that this should lead to contractions, which is
consistent with experimental findings \cite{ennomani2016architecture}. 

After having discussed the effects of motor stepping on the active pressure in
systems with $\mathbf P = 0$, we ask how the situation changes in polar systems.
In polar system an additional contribution, $-(\sigma_{01} -
\frac{\gamma_1}{\gamma_0}\sigma_0)|\mathbf P|^2$, exists. For symmetric motors,
where $\sigma_{01}=\sigma_{10}$ this implies that the active pressure generated
by a network of symmetric motors and passive crosslinks is
strongest in apolar regions of the system and subsides in polar regions, since the
polar and apolar contributions to the active stress appear in
Eq.~{(\ref{eq:stress_final})} with opposite signs. We plot the magnitude of the
active pressure $\Pi^\mathrm{(A)} \simeq 1 -|\mathbf P|^2$ as a function of
$|\mathbf P|$ in Table \ref{tab:motor_to_effects}. This is reminiscent of the
behavior predicted in the frameworks of a sparsely crosslinked system in
\cite{gao2015multiscalepolar}. In contrast the effects of non-symmetric motors can be enhanced in
polar regions. Consider again, the example of a motor with one static head
that preferentially binds near one of the filament ends and a mobile head that
acts uniformly. For this motor
$\sigma_{10} = \gamma_1 V_0$ and $\sigma_{01} = 0$ and $\sigma_0 = \gamma_0
V_0/2$. It is thus predicted to generate twice the amount of active pressure in
a polar network than in an apolar one and $\Pi^\mathrm{(A)}\simeq (1+|\mathbf P|)/2$, see the
table \ref{tab:motor_to_effects} for a plot of the active pressure
$\Pi^\mathrm{(A)}$ as a function of $|\mathbf P|$.  This is reminiscent of the
motor dynein in spindles, which is though to generate the most prominent
contractions near the spindle poles, which are polar
\cite{brugues2012nucleation}.

Finally, we ask how filament length affects the active pressure. Looking at the
definitions of the nematic and polar activity Eqs.~{(\ref{eq:AQ}, \ref{eq:App})}
and remembering the definition and scaling of the coefficient in there (see
App.~{(\ref{app:coeffcients})}), we notice that the active pressure scales as
$L^4$. Since the viscosity scaled
with $L^2$, this predicts that systems with shorter filaments
contract slower than systems with longer filaments. This effect has observed for
dynein based contractions in-vitro \cite{foster2017connecting}.

\subsection{Tuning axial stresses, buckling and aster formation}
 Motors in active filament networks generate anisotropic (axial) contributions
 to the stress, which  can lead to large scale instabilities in materials
 with nematic order \cite{simha2002hydrodynamic, kruse2005generic,
 saintillan2008instabilitiesPRL, furthauer2012taylor}. At larger active
 stresses, nematics are unstable to splay deformations in systems that are
 contractile along the nematic axis, and to bend deformations in systems that
 are extensile along the nematic axis \cite{kruse2005generic,
 marchetti2013hydrodynamics}. In both cases, the instabilities set in when the
 square root of ratio of the elastic (bend or splay) modulus that opposes the
 deformation to the active stress - also called the Fr\'eedericksz length -
 becomes comparable to the systems size. We now discuss which motor properties
 control the emergence of these instabilities, and how a system can be tuned
 exhibit bend or splay deformations. For this we ask how axial stresses, which
 are governed by  the activity parameters $\mathcal A^{(\mathcal Q)}$ and
 $\mathcal{A}^{(\mathbf P)}$, are set in our system.

The magnitude $S$ of the axial stress along the nematic axis is given by
\begin{equation}
    S = -\rho^2 q\left( \mathcal A^{(\mathcal Q)} -  \mathcal A^{(\mathbf
    P)}|\mathbf P|^2\right),
    \label{eq:axial_mag}
\end{equation}
where we defined the nematic order parameter $q$, as the largest eigenvalue of
${\mathcal Q} - \mathrm{Tr}(\mathcal Q)\mathcal{I}/3$; see
Eq.~{(\ref{eq:stress_final})}. The axial stress is
contractile along the nematic axis if $S$ is positive and extensile if $S$ is
negative. Comparing Eqs.~{(\ref{eq:axial_mag}, \ref{eq:mpress})} we find that
$S = q(\Pi^\mathrm{(A)} + \rho^2\alpha K_0)$ and in the limit where $K_0\to 0$,
where motor elasticity is negligible, $S = q\Pi^\mathrm{(A)}$. We discussed how
$\Pi^\mathrm{(A)}$ is set for different types of crosslinks in the previous
section; see Table \ref{tab:motor_to_effects}. 

The prototypical active nematic \cite{sanchez2012spontaneous} which consists
of apolar bundles of microtubules actuated by the kinesin motors and is axial extensile. 
In our theory, an axial extensile stress (i.e. $S<0$) in an apolar
system ($\mathbf P = 0$) implies that $\mathcal A^{(\mathcal Q)} = \sigma_{10}
-\sigma_0\frac{\gamma_1}{\gamma_0} +\frac{L^2}{12}K_0 >0$. 
This can be achieved either by crosslinks that act uniformly (i.e. $\sigma_{10}
-\sigma_0\frac{\gamma_1}{\gamma_0} = 0$) and generate a spring like response
that induces $K_0>0$ or by crosslinks that have non-uniform motor stepping behavior
which generates $\sigma_{10} -\sigma_0\frac{\gamma_1}{\gamma_0}>0$. The latter
implies either a non-symmetric motor crosslinks, or the presence of more than
one kind of crosslinks, as was discussed more extensively earlier in the context
of active pressure. At high enough active stress we expect systems with negative
$S$ to become unstable towards buckling. This has been observed in
\cite{senoussi2019tunable,strubing2019wrinkling}.

Conversely axial contractile behavior can be achieved if either $K_0<0$ or
$\sigma_{10} -\sigma_0\frac{\gamma_1}{\gamma_0}<0$. At high enough active
stress, such systems can become unstable towards an aster forming transition, as
seen in \cite{foster2015active}. 

Note that $S\simeq\Pi^\mathrm{(A)} + \rho^2\alpha K_0$, implies that $S$ and
$\Pi^\mathrm{(A)}$ need not be the same if $K_0\neq 0$. In particular when
$\Pi^\mathrm{(A)}$ and $K_0$ have opposite signs systems can exist, which are
axially extensile while being bulk contractile and vice versa. 

We finally note that the magnitude of axial stresses changes if the system
transitions from apolar to polar, if the origin of the axial stresses is motor
stepping but not if the origin of the axial stresses is the effective spring
like behavior of motor, since
$\mathcal A^{(\mathcal Q)}$, but not  $\mathcal A^{(\mathbf P)}$, depends on
$K_0$, see Eqs.~{(\ref{eq:App}, \ref{eq:AQ})}. In systems in which the active
stress is generated by the stepping of symmetric motor-crosslink, $|S|$ is
highest nematic apolar phase ($|\mathbf P|=0$), while systems made from
non-symmetric crosslinks generate the most stress when polar ($|\mathbf P|=1$);
see Table \ref{tab:motor_to_effects}. This opens the possibility that a system
can overcome the threshold towards an instability when its other dynamics drives
it from nematic apolar to polar arrangements or vice versa. We suggest that the
buckling instabilities discussed in
\cite{senoussi2019tunable,strubing2019wrinkling} should be interpreted in this
light.

\section{Discussion}
\label{sse:discussion}
 In this paper, we asked how the properties of motorized crosslinkers that act
 between the filaments of a highly crosslinked polymer network set the large
 scale properties of the material. 
 
 For this, we first develop a method for quantitatively stating what the
 properties of motorized crosslinks are. We introduce a generic phenomenological
 model for the forces that crosslink populations exert between the filaments
 which they connect; see Eq.~{(\ref{eq:generic_force_density})}. This model
 describes forces that are (i) proportional to the distance ($K$), and (ii) the
 relative rate of displacement ($\gamma$). Finally
 (iii) it describes the active motor forces ($\sigma$) that crosslinks can
 exert. Importantly, forces from crosslinkers ($K, \gamma, \sigma$) can
 depend on the position on the two filaments which they couple.
 This allows the description of a wide range of motor properties, such as
 end-binding affinity, end-dwelling, and even the description of non-symmetric
 crosslinks that consist of motors with two heads of different properties. 

 We next derived the stresses and forces generated on large time and length scale,
 given our phenomenological crosslink model. We find that the emergent material
 stresses depend only on a small set of moments; see
 Eq.~{(\ref{eq:moment_definition})} of the crosslink properties. 
 These moments
 are effectively descriptions of the expectation value of the force exerted
 between two filaments given their positions and relative orientations. 
 The resulting stresses, forces, and filament reorientation rates
 (Eqs.~{(\ref{eq:stress_final}, \ref{eq:v_final}, \ref{eq:kinetic_angular})})
 recover the symmetries and structure predicted by phenomenological theories for
 active materials, but beyond that provide a way of identifying how specific
 micro-scale processes set specific properties of the material. 

 We discussed how four key aspects of the dynamics of highly crosslinked
 filament networks can be tuned by the micro-scale properties of motors and
 filaments. In particular we discussed how  (i) the highly anisotropic viscosity
 of the material is set; (ii) how active self-straining is regulated; (iii) how
 contractile or extensile active pressure can be generated; (iv) which motor
 properties regulate the axial active nematic and bipolar stresses, which can
 lead to large scale instabilities. 

 Our theory makes specific predictions for the effects of distinct classes of
 crosslinkers on cytoskeletal networks. Intriguingly these predictions suggest
 explanations for phenomena experimentally seen, but currently poorly
 understood.  

 Experiments have shown that mixtures of actin filaments and myosin
 molecular motors can spontaneously contract, but only in the presence of an
 additional passive crosslinker \cite{ennomani2016architecture}.  Our theory
 allows us to speculate on explanations for this observation.  In the crosslink
 classification that we introduced, myosin, which
 form large mini-filaments, is a symmetric motor crosslink; see
 Fig.~{(\ref{fig:motor_types})}.  We find that symmetric motor crosslinks, which
 have two heads that act the same can generate contractions only in the presence
 of an additional crosslinker that helps break the balance between
 $\gamma_1/\gamma_0\sigma_0$ and $\sigma_{01}$ in the active pressure; see
 Eq.~{(\ref{eq:pi_motors_only_nema})} and Table~\ref{tab:motor_to_effects}.
 Further work will be needed to explore whether this connection can be made
 quantitative.

 A second observation that was poorly understood prior to this work is the sliding
 motion of microtubules in meiotic Xenopus spindles, which are the structures
 which segregate chromosomes during the meiotic cell division. These spindles
 consist of inter-penetrating arrays of  anti-parallel microtubules, which are
 nematic near the chromosomes, and highly polar near the spindle poles. In most
 of the spindle the two anti-parallel populations of microtubules slide past
 each other, at near constant speed driven by the molecular motor Eg-5 Kinesin,
 regardless of the local network polarity. Our earlier work
 \cite{furthauer2019self} showed that active self straining explains this
 polarity independent motion. The theory that we develop here provides the tools
 to explore the behavior of different motors and motor mixtures which will allow
 us to investigate the mechanism by which different motors in the spindle shape
 its morphology. This will help to explain complex behaviors of spindles such as
 the barreling instability \cite{oriola2020active} that gives spindles their
 characteristic shape or the observation that spindles can fuse
 \cite{gatlin2009spindle}.

 Our theory provides specific predictions on how changing motor properties can
 change the properties of the material which they constitute, it can enable the
 design of new active materials. We predict the expected large scale properties
 of a material, in which an experimentalist had introduced engineered crosslinks
 with controlled properties. With current technology, an experimentalist could
 engineer a motor that preferentially attaches one of its heads to a specified
 location on a filament, while its walking head reaches out into the network.
 Or, as has already been demonstrated in studies by the Surrey Lab
 \cite{roostalu2018determinants} the difference in the rates of filament growth
 and motor walking speeds, could be exploited to generate different dynamic
 motor distributions on filaments. This design space will provide ample room to
 experimentally test our predictions, and use them to engineer systems with
 desirable properties. Finally recent advances in optical control of motor
 systems \cite{ross2019controlling} could be used to provide spatial control.

 The theory presented here does however make some simplifications.
 Importantly, we neglected that the distribution of bound crosslinks on
 filaments themselves in general depends on the configuration of the network.
 This means that the crosslink moments can themselves be functions of the local
 network order parameters. Effects like this have been argued to be important
 for instance when explaining the transition from contractile to extensile
 stresses in ordering microtubule networks \cite{lenz2020reversal} and the
 physics of active bundles \cite{kruse2003self}. Such effects
 can be recovered when making the interactions $K, \gamma, \sigma$ in the
 phenomenological crosslink force model Eq.~{(\ref{eq:generic_force_density})}
 functions of $\mathbf p_i$, and $\mathbf p_j$. This will be the topic of a
 subsequent publication. 

 In summary, in this paper we derived a continuum theory for systems made from
 cytoskeletal filaments and motors in the highly crosslinked regime. Our theory
 makes testable predictions on the behavior of the emerging system, provides a
 unifying framework in which dense cytoskeletal systems can be understood from
 the ground up, and provides the design paradigms, which will enable the creation
 of active matter systems with desirable properties in the lab.

 {\bf Acknowledgements} We thank Meridith Betterton and Adam Lamson for
 insightful discussions. We also thank Peter J. Foster and James F. Pelletier for
 feedback on the manuscript.  DN acknowledges support by the National Science Foundation
 under awards DMR-2004380 and DMR-0820484.  MJS acknowledges support by the
 National Science Foundation under awards DMR-1420073 (NYU MRSEC), DMS-1620331,
 and DMR-2004469.


\newpage
\newpage
\begin{appendix}
\section{Detailed derivation of the equations of motion}\label{app:calculate}
In the following we derive the equations of motion for the highly crosslinked
active network. We start by using Eq.~{(\ref{eq:low_order_vi},
\ref{eq:rod_torque})} and obtain
\begin{eqnarray}
    \dot{\mathbf p}_i &=& \left( \mathcal{I} -\mathbf p_i \mathbf p_i
    \right)\cdot\left\{\mathbf p_i\cdot(\mathcal U +\frac{12}{\gamma_0 L^2}\frac{\mathcal E}{
        \rho^2})
+\frac{12}{\gamma_0L^2}   A^{(\mathbf P)}\mathbf P \right\}
\nonumber\\
&-&\frac{1}{\rho} \bar{\mathbf T}_i^\mathrm{(drag)},
    \label{eq:kinetic_angular_supp}
\end{eqnarray}
The torque due to drag with the medium is
\begin{eqnarray}
    &&\bar{\mathbf T}^{(drag)}_i = {\mathbf T}^{(drag)}_i 
    \nonumber\\
    &+&\left( \mathcal{I} -\mathbf p_i\mathbf p_i \right)\cdot\left( 
    \frac{12}{\gamma_0 L^2} +\mathbf p_i\cdot\frac{\nabla\rho}{\rho}\right)
    \left(\mathbf F_i^\mathrm{(drag)} -\frac{1}{\rho}\mathbf f \right).
    \nonumber\\
\end{eqnarray}
This implies
\begin{eqnarray}
    \mathbf\omega &=& \mathbf P\cdot(\mathcal U +\frac{12}{\gamma_0 L^2}\frac{\mathcal E}{
        \rho^2}) - \mathcal T : (\mathcal U +\frac{12}{\gamma_0 L^2}\frac{\mathcal E}{
        \rho^2})
    \nonumber\\
    &+& \frac{12}{L^2}A^{(\mathbf P)}\left( \mathbf P -\mathcal Q\cdot
        \mathbf P \right)
    -\frac{1}{\rho}\omega^\mathrm{(drag)},
    \nonumber\\
\end{eqnarray}
where
\begin{equation}
    \omega^\mathrm{(drag)} = \left<\bar{\mathbf T}^{(drag)}_i\right>
\end{equation}
and
\begin{eqnarray}
\mathcal H &=& \mathcal Q\cdot(\mathcal U +\frac{12}{\gamma_0 L^2}\frac{\mathcal E}{
        \rho^2}) - \mathcal S:(\mathcal U +\frac{12}{\gamma_0 L^2}\frac{\mathcal E}{
        \rho^2})    
\nonumber\\
&+&\frac{12}{\gamma_0 L^2}A^{(\mathbf P)}\left(
        \mathbf P\mathbf P -\mathcal T \cdot\mathbf P \right)
        \nonumber\\
    &-&\frac{1}{\rho}\mathcal H^\mathrm{(drag)},
\end{eqnarray}
where 
\begin{equation}
    \mathcal{H}^\mathrm{(drag)} =\left<\mathbf p _i\bar{\mathbf
    T}^{(drag)}_i\right>.
\end{equation}
Furthermore we note that
\begin{equation}
    \mathbf j = \frac{\sigma_0}{\gamma_0}(\mathbf P \mathbf P - \mathcal
    Q)+\frac{1}{\gamma_0\rho} \mathbf j^\mathrm{(drag)} +\mathcal{O}\left( L^2
    \right),
\end{equation}
and
\begin{equation}
    \mathcal J = \frac{\sigma_0}{\gamma_0}(\mathcal Q \mathbf P - \mathcal
    T)+\frac{1}{\gamma_0\rho} \mathcal J^\mathrm{(drag)} +\mathcal{O}\left( L^2
    \right),
\end{equation}
where
\begin{equation}
    \mathbf j^\mathrm{(drag)} = -\frac{1}{\gamma_0}\left< \mathbf p_i 
    \left(\mathbf F_i^\mathrm{(drag)} -\frac{1}{\rho}\mathbf f \right)\right>
\end{equation}
and
\begin{equation}
    \mathcal J^\mathrm{(drag)} = -\frac{1}{\gamma_0}\left< \mathbf p_i \mathbf p_i 
    \left(\mathbf F_i^\mathrm{(drag)} -\frac{1}{\rho}\mathbf f \right)\right>.
\end{equation}
Putting all of this together, we arrive at an expression for the networks stress in terms of the current
distribution of filaments,
\begin{eqnarray}
    \mathbf\Sigma &=&
    -\rho^2\left(\chi:\mathcal U  +\alpha K_0\mathcal{I}\right)
    \nonumber\\
    &-&\rho^2\left( 
        A^{(\mathcal Q)}\mathcal Q
    -A^{(\mathbf P)}\mathcal T\cdot\mathbf P\right) 
    \nonumber\\
    &-& \bar\chi:\mathcal E
    +\bar{\mathbf\Sigma} + \rho \mathbf\Sigma^\mathrm{(drag)}.
    \nonumber\\
\end{eqnarray}
where
\begin{equation}
    \mathbf \Sigma^\mathrm{(drag)} = -\gamma_0\frac{L^2}{12}\mathcal
    H^\mathrm{(drag)} -\gamma_1\mathbf j^\mathrm{(drag)}
\end{equation}
and at a similar equation for the motion of filament $i$
\begin{eqnarray}
    \mathbf v_i -\mathbf v &=& -\frac{\sigma_0}{\gamma_0}\left( \mathbf p_i -\mathbf
    P\right) 
    \nonumber\\
    &-&\frac{\gamma_1}{\gamma_0}\left(
        \begin{array}{c}
        (\mathbf p_i-\mathbf P)\cdot(\mathcal U +\frac{12}{\gamma_0 L^2}\frac{\mathcal E}{
        \rho^2})\\ 
    -(\mathbf p_i\mathbf p_i\mathbf p_i-\mathcal T):(\mathcal U
+\frac{12}{\gamma_0 L^2}\frac{\mathcal E}{ \rho^2})
\end{array}
\right)
    \nonumber\\
    &+&\frac{1}{\gamma_0}\frac{12\gamma_1}{L^2\gamma_0}A^{(\mathbf P)}\left( \mathbf p_i\mathbf
    p_i-\mathcal Q \right)\cdot \mathbf P
    \nonumber\\
    &-&\frac{\nabla\rho}{\gamma_0\rho}\cdot\left[
        \begin{array}{c}
        A^{\mathcal Q} \left( \mathbf
            p_i \mathbf p_i - \mathcal Q
        \right) 
        \\- A^{(\mathbf P)}\left( \mathbf p_i \mathbf P + \mathbf
        P \mathbf p_i -2\mathbf P\mathbf P\right)
    \end{array}
    \right]
        \nonumber\\
    &-&\left( \mathbf p_i\mathbf p_i -\mathcal Q):\nabla\frac{\mathcal E}{\rho} \right)
    -\frac{1}{\rho} \mathbf v^\mathrm{(drag)}
\end{eqnarray}
where 
\begin{equation}
    \frac{1}{\rho}\mathbf v^\mathrm{(drag)} = \frac{1}{\rho\gamma_0}\mathbf
    F_i^\mathrm{(drag)} + \mathcal O\left(1/\rho^2\right).
\end{equation}

\section{Crosslink Moments}
\label{app:coeffcients}
The crosslink moment which enter the hydrodynamic descriptions are defined from
moments of crosslinker mediated filament-filament forces. Specifically,
\begin{eqnarray}
    K_0  &=& \left\lfloor K(s_i, s_j)\right\rceil_{\Omega(\mathbf x_i)}^{ij}, 
    \\
    K_1  &=& \left\lfloor s_i K(s_i, s_j)\right\rceil_{\Omega(\mathbf x_i)}^{ij}, 
    \\
    \gamma_0  &=& \left\lfloor \gamma(s_i, s_j)\right\rceil_{\Omega(\mathbf x_i)}^{ij}, 
    \\
    \gamma_1  &=& \left\lfloor s_i \gamma(s_i, s_j)\right\rceil_{\Omega(\mathbf
    x_i)}^{ij}, 
    \\
    \sigma_0  &=& \left\lfloor\sigma(s_i, s_j)\right\rceil_{\Omega(\mathbf x_i)}^{ij}, 
    \\
    \sigma_{10}  &=& \left\lfloor s_i \sigma(s_i,
    s_j)\right\rceil_{\Omega(\mathbf x_i)}^{ij}, 
\end{eqnarray}
and
\begin{eqnarray}
    \sigma_{01}  &=& \left\lfloor s_j \sigma(s_i,
    s_j)\right\rceil_{\Omega(\mathbf x_i)}^{ij}. 
\end{eqnarray}
\section{Angular Momentum Fluxes and antisymmetric stresses}\label{app:angular}
The spin and orbital angular momenta obey the continuity equations
\begin{equation}
    \dot{\mathbf\ell}_i = \sum_j\mathbf T_{ij} + \mathbf  T_{i}^\mathrm{(drag)}
\end{equation}
and
\begin{equation}
\mathbf x_i \times \dot{\mathbf g}_i  = \sum_j\mathbf x_i\times \mathbf
F_{ij} + \mathbf x_i \times \mathbf  F_{i}^\mathrm{(drag)},
\end{equation}
where we used Eq.{~(\ref{eq:Newton_single_particle})} and that $\dot{\mathbf
x}_i$ is parallel to $\mathbf g$.
We and introduce
the densities of spin and orbital angular momentum which are
\begin{equation}
    \mathbf \ell = \sum_i\delta(\mathbf x-\mathbf x_i)\ell_i,
    \label{eq:spin_angular_density_definition}
\end{equation}
and
\begin{equation}
    \mathbf \ell^\mathrm{(orb)} = \sum_i\delta(\mathbf x-\mathbf
    x_i)\mathbf x_i\times\mathbf g_i,
    \label{eq:orbital_angular_density_definition}
\end{equation}
respectively. They obey continuity equations
\begin{eqnarray}
\partial_t\mathbf\ell 
+ \nabla\cdot\sum_i\left(\delta(\mathbf x -
\mathbf x_i)\mathbf v_i  \mathbf
\ell_i\right)
=
\sum_{i,j}\delta(\mathbf x -\mathbf x_i) \mathbf T_{ij}+\mathbf \tau,
\nonumber\\
\label{eq:spin_continuity}
\end{eqnarray}
where
\begin{equation}
    \mathbf \tau = \sum_i \delta(\mathbf x- \mathbf x_i) \mathbf
    T_i^\mathrm{(drag)}
    \label{eq:define_spin_torque}
\end{equation}
and
\begin{eqnarray}
\partial_t\mathbf\ell^\mathrm{(orb)} 
+ \nabla\cdot\sum_i\delta(\mathbf x -
\mathbf x_i)\mathbf v_i  
\mathbf x_i\times\mathbf g_i
= \nonumber\\
\sum_{i,j}\delta(\mathbf x -\mathbf x_i)\mathbf x_i\times \mathbf
F_{ij} + \mathbf x\times \mathbf f.
\label{eq:orb_continuity}
\end{eqnarray}

The first term on the right hand side of Eq.~{(\ref{eq:orb_continuity})}
describes the orbital angular momentum transfer by crosslink interactions. It can be
rewritten as the sum of an orbital angular momentum flux $\mathcal
M^\mathrm{(orb)}$ and a source term related to the antisymmetric part of the
stress tensor $\mathbf\Sigma$,
\begin{eqnarray}
&~&\sum_{i,j}\delta(\mathbf x -\mathbf x_i)\mathbf x_i\times \mathbf
F_{ij}
\nonumber\\
&=&
\sum_{i,j}\delta\left(\mathbf x -\mathbf x_i\right)\frac{\mathbf x_i+\mathbf
x_j}2\times \mathbf F_{ij}
\nonumber\\
&&+ \sum_{i,j}\delta\left(\mathbf x -\mathbf x_i\right)
\frac{\mathbf x_i-\mathbf x_j}2\times \mathbf F_{ij}
\nonumber\\
&=&
\nabla\cdot \mathcal{M}^\mathrm{(orb)} + 2 \mathbf\sigma^a + \mathcal{O}(\mathbf d_{ij}^3),
\end{eqnarray}
where the orbital angular momentum flux is
\begin{equation}
\mathcal{M}^\mathrm{(orb)} = -\sum_{i,j}\delta\left(\mathbf x -
\mathbf x_i\right)\frac{\mathbf x_i-\mathbf
x_j}2\left(\frac{\mathbf x_i+\mathbf
x_j}2\times \mathbf F_{ij}\right)
\end{equation}
and  
\begin{equation}
\mathbf\sigma^a
=
\sum_{i,j}\delta\left(\mathbf x -\mathbf x_i\right)
\frac{\mathbf x_{i}-\mathbf x_j}4\times \mathbf F_{ij},
\end{equation}
which is the pseudo-vector notation for the antisymmetric part of the stress
$\mathbf\Sigma$ such that in index notation, 
\begin{equation}
\mathbf\sigma^a_\alpha = 
\frac{1}{2}\epsilon_{\alpha\beta\gamma}\mathbf\Sigma_{\beta\gamma},
\end{equation}
where used the Levi-Civita symbol $\varepsilon_{\alpha\beta\gamma}$ and
summation over repeated greek indices is implied.

Similarly, the first term on the right hand side of Eq.~{(\ref{eq:spin_continuity})}
describes the spin angular momentum transfer by crosslink interactions. It can be
rewritten as the sum of an orbital angular momentum flux $\mathcal
M$ and a source term related to the antisymmetric part of the
stress tensor $\mathbf\Sigma$,
\begin{eqnarray}
&~&\sum_{i,j}\delta(\mathbf x -\mathbf x_i) \mathbf
T_{ij}
\nonumber\\
&=&
\sum_{i,j}\delta\left(\mathbf x -\mathbf x_i\right)\left(\mathbf T_{ij} + \frac{\mathbf x_i-\mathbf
x_j}2\times \mathbf F_{ij}\right)
\nonumber\\
&& - \sum_{i,j}\delta\left(\mathbf x -\mathbf x_i\right)
\frac{\mathbf x_i-\mathbf x_j}2\times \mathbf F_{ij}
\nonumber\\
&=&
\nabla\cdot \mathcal{M} - 2 \mathbf\sigma^a + \mathcal{O}(\mathbf d_{ij}^3),
\end{eqnarray}
where the spin angular momentum flux
\begin{equation}
    \mathcal M = -\sum_{i,j}\delta\left(\mathbf x -\mathbf x_i\right)\frac{\mathbf x_i-\mathbf
x_j}2\left(\mathbf T_{ij} + \frac{\mathbf x_i-\mathbf
x_j}2\times \mathbf F_{ij}\right)
    \label{eq:spin_momentum_density_flux}
\end{equation}

After defining the total
and spin angular momentum fluxes as
\begin{eqnarray}
    \mathcal{M}^\mathrm{(tot)} &=& \mathcal M +\mathcal M^\mathrm{(orb)}
    \nonumber\\
    &=& -\sum_{i,j}\delta\left(\mathbf x -
    \mathbf x_i \right)\frac{\mathbf x_i-\mathbf x_j}2\left( \mathbf T_{ij} + \mathbf x_i\times\mathbf F_{ij} \right),
    \nonumber\\
\end{eqnarray}
we finally write down the statements of 
angular momentum conservation
\begin{equation}
\nabla\cdot\mathcal{M}^\mathrm{(tot)} + \mathbf x\times \mathbf f +\mathbf \tau
= 0,
\label{eq:angular_momentum_continuum}
\end{equation}
spin angular momentum continuity
\begin{equation}
\nabla\cdot\mathcal{M} - 2\mathbf\sigma^a + \mathbf \tau = 0,
\label{eq:spin_angular_momentum_continuum}
\end{equation}
and orbital angular momentum continuity
\begin{equation}
\nabla\cdot\mathcal{M}^\mathrm{(orb)} + 2\mathbf\sigma^a + \mathbf x\times
\mathbf f = 0,
\label{eq:orbital_angular_momentum_continuum}
\end{equation}
where we dropped inertial terms. We note that the antisymmetric stress
$\mathbf\Sigma^a$ acts to transfer spin to orbital angular momentum.
Importantly, the total angular momentum is conserved as evident from the form of
Eq.~{(\ref{eq:angular_momentum_continuum})}.

\section{The Ericksen Stress}
\label{app:steric}
In this appendix we derive the effects of steric interactions on the system.
As stated in the main text, steric interactions are best described in terms of a
potential $e(\mathbf x_i, \mathbf p_i)$, which depends on all particle positions
and orientations. The steric free energy of the system is $E = \int_{\mathcal V} e d^3x$ 
where $\mathcal V$ is the volume of the system. 
For the treatment to follow we shall assume the steric interactions do not
depend on the polar, but only on the nematic order of the system.
Then a generic variation of the
systems free energy can be written as
\begin{eqnarray}
    \delta E &=& \int_{\partial\mathcal V} \left(  e u_\gamma
+\frac{\partial e}{\partial(\partial_\gamma Q_{\alpha\beta})}\delta
Q_{\alpha\beta} \right)dS_\gamma
\nonumber\\
&-&\int_{\partial\mathcal V}\left(\mu \delta \rho +\mathcal E_{\alpha\beta}
\delta Q_{\alpha\beta}\right)
\end{eqnarray}
where we defined the chemical potential
\begin{equation}
    \mu = -\frac{\partial e}{\partial \rho}
\end{equation}
and the distortion field
\begin{equation}
    \mathcal E_{\alpha\beta} = -\frac{\partial e}{\partial
    Q_{\alpha\beta}}
    +\partial_\gamma\frac{\partial e}{\partial (\partial_\gamma
    Q_{\alpha\beta})},
\end{equation}
and introduced the infinitesimal deformation field $\mathbf u$.
Now, any physically well defined free energy density needs to obey translation
invariance. Thus $\delta E = 0 $ for any pure translation, which is the
transformation where $\delta \rho = - u_\gamma\partial_\gamma\rho$,  
$\delta Q_{\alpha\beta} = - u_\gamma\partial_\gamma Q_{\alpha\beta}$, $u_\gamma$
is a constant. Thus
\begin{eqnarray}
    &&\partial_\beta\left(\left(e + Q_{\mu\nu}\mathcal E_{\mu\nu} +\mu\rho \right)\delta_{\alpha\beta}  
-\frac{\partial e}{\partial(\partial_\beta Q_{\gamma\mu})}\partial_\alpha
Q_{\gamma\mu} \right)
\nonumber\\
&=&\rho \partial_\alpha \mu + Q_{\mu\nu}\partial_\alpha \mathcal
E_{\mu\nu},
\end{eqnarray}
which is the Gibbs-Duhem relation used in the main text, where
\begin{equation}
    \bar \Sigma_{\alpha\beta} = \left(e + Q_{\mu\nu}\mathcal E_{\mu\nu} +\mu\rho
    \right)\delta_{\alpha\beta}  -\frac{\partial e}{\partial(\partial_\beta
    Q_{\gamma\mu})}\partial_\alpha Q_{\gamma\mu}.
\end{equation}

\end{appendix}

\end{document}